\documentclass[fleqn,usenatbib]{mnras}
\usepackage[utf8]{inputenc}
\usepackage[T1]{fontenc}
\usepackage{ae,aecompl}
\usepackage{flushend}

\usepackage{amsmath,amssymb,leftidx}
\usepackage{cleveref,graphicx}
\usepackage{algorithm,algorithmicx,algpseudocode}

\usepackage[dvipsnames]{xcolor}
\usepackage{tikz}
\usetikzlibrary{calc}

\usepackage{siunitx}
\DeclareSIUnit{\year}{yr}

\usepackage{newtxtext,newtxmath}

\hypersetup{pdfauthor={B. V. Lehmann et al.},
   pdftitle={Model-independent discovery prospects for primordial black holes at LIGO},
   pdfkeywords={gravitational waves, methods: analytical, methods: numerical, dark matter, black hole mergers},
   bookmarksnumbered=true,
   linkcolor=blue,
   citecolor=blue,
   filecolor=cyan,
   urlcolor=blue}

\newcommand\DP{{DP}}
\newcommand\OmegaPBH{\Omega_{\mathrm{PBH}}}

\newcommand\OmegaDM{\Omega_{\mathrm{DM}}}
\newcommand\OmegaGW{\Omega_{\mathrm{GW}}}

\newcommand\rhoeq{\rho_{\mathrm{eq}}}
\newcommand\fmaxmono{f_{\mathrm{max}}}
\newcommand\fpbh{f_{\mathrm{PBH}}}
\newcommand\mchirp{M_c}
\newcommand\mdpmin{m_{\mathrm{\DP}}^{\mathrm{min}}}
\newcommand\mdpmax{m_{\mathrm{\DP}}^{\mathrm{max}}}
\newcommand\rdp{r_{\mathrm{\DP}}}
\newcommand\fdp{f_{\mathrm{\DP}}}
\newcommand\du{\mathrm{d}}
\newcommand\dd{\,\du}
\newcommand\algfunc[1]{\operatorname{\texttt{#1}}}
\newcommand{\DPBH}{\DP BH}
\newcommand{\bb}[1]{\vec{\mathrm{#1}}}
\newcommand{\cvec}{\bb{\mathcal C}}
\newcommand{\constraintlabels}{The labelled constraints are from BH evaporation \protect\cite[\texttt{evap};][]{Carr:2009jm}, Hyper Suprime-Cam \protect\cite[\texttt{HSC};][]{Niikura:2017zjd,Smyth:2019whb}, Kepler \protect\cite[\texttt{K};][]{Griest:2013aaa}, OGLE \protect\cite[\texttt{Ogle};][]{Wyrzykowski:2011tr}, EROS-II \protect\cite[\texttt{EROS};][]{Tisserand:2006zx}, MACHO \protect\cite[\texttt{M};][]{Allsman:2000kg}, and CMB observables \protect\cite[\texttt{CMB};][]{Ali-Haimoud:2016mbv,Carr:2017jsz}. Other constraints may also apply, but their inclusion does not influence our qualitative conclusions (see \cref{sec:optimization}).}

\title[Discovery prospects for PBH at LIGO]{Model-independent discovery prospects for primordial black holes at LIGO}

\author[B. V. Lehmann, S. Profumo and J. Yant]{Benjamin V. Lehmann,$^{1,2}$\thanks{E-mail: \href{mailto:benvlehmann@gmail.com}{benvlehmann@gmail.com}}
Stefano Profumo,$^{1,2}$\thanks{E-mail: \href{mailto:profumo@ucsc.edu}{profumo@ucsc.edu}}
and Jackson Yant$^{1,2,3}$\thanks{E-mail: \href{mailto:jackson.yant.phys@gmail.com}{jackson.yant.phys@gmail.com}}
\\
$^{1}$Department of Physics, University of California Santa Cruz, 1156 High St, Santa Cruz, CA 95064, USA\\
$^{2}$Santa Cruz Institute for Particle Physics, 1156 High St, Santa Cruz, CA 95064, USA\\
$^{3}$Department of Physics and Astronomy, Dartmouth College, Hanover, NH 03755, USA}

\date{Accepted XXX. Received YYY; in original form ZZZ}
\pubyear{2020}

\begin{document}
\label{firstpage}
\pagerange{\pageref{firstpage}--\pageref{lastpage}}
\maketitle

\begin{abstract}
    Primordial black holes may encode the conditions of the early Universe, and may even constitute a significant fraction of cosmological dark matter. Their existence has yet to be established. However, black holes with masses below $\sim\SI{1}{M_\odot}$ cannot form as an endpoint of stellar evolution, so the detection of even one such object would be a smoking gun for new physics, and would constitute evidence that at least a fraction of the dark matter consists of primordial black holes. Gravitational wave detectors are capable of making a definitive discovery of this kind by detecting mergers of light black holes. But since the merger rate depends strongly on the shape of the black hole mass function, it is difficult to determine the potential for discovery or constraint as a function of the overall abundance of black holes. Here, we directly maximize and minimize the merger rate to connect observational results to the actual abundance of observable objects. We show that LIGO can discover mergers of light primordial black holes within the next decade even if such black holes constitute only a very small fraction of dark matter. A single merger event involving such an object would (i) provide conclusive evidence of new physics, (ii) establish the nature of some fraction of dark matter, and (iii) probe cosmological history at scales far beyond those observable today.
\end{abstract}


\section{Introduction}
\label{sec:introduction}
The detection of black hole binaries with LIGO \citep{Abbott:2016blz,Abbott:2016nmj,Abbott:2017vtc,Abbott:2017oio,Abbott:2017gyy,LIGOScientific:2018mvr,Abbott:2020uma,LIGOScientific:2020stg} has heralded a new era in probing the nature and behavior of compact objects in our Universe. In the past several years, gravitational wave detectors have directly confirmed the existence of black holes \citep{TheLIGOScientific:2016pea}, provided powerful tests of general relativity \citep{TheLIGOScientific:2016src}, and ushered in the era of multimessenger astronomy \citep{GBM:2017lvd,Monitor:2017mdv}. But as gravitational wave observatories continue to probe the black hole population, they are poised to make yet another significant discovery: mergers may provide direct evidence for the existence of primordial black holes (PBH).

PBH may form in the early Universe without stellar progenitors, and they have been intensely studied as potential probes of cosmology and high-energy physics \citep[see e.g.][]{Carr:2003bj,Khlopov:2008qy,Calmet:2014dea}. In the simplest scenario, PBH are formed by the gravitational collapse of large density perturbations on small scales \citep{1967SvA....10..602Z,10.1093/mnras/152.1.75}. Other formation channels, such as gravitational collapse in a dark sector \citep{Shandera:2018xkn}, or the collapse of another compact object due to new physics \citep[see e.g.][]{deLavallaz:2010wp}, can also result in observable black holes of non-stellar origin. Since the interactions of massive PBH are dominated by gravitation, and since a collection of black holes is fluid-like on sufficiently large scales, PBH are a natural candidate for cosmological dark matter, a possibility that was recognized shortly after their existence was postulated \citep{Chapline:1975ojl}.

In the decades since, the PBH population has been constrained by various astrophysical and cosmological means, but never fully ruled out as a dark matter candidate \citep{Carr:2009jm,Carr:2016drx,Carr:2020gox,Lehmann:2018ejc,Carr:2020xqk}. Soon after the first observation of a binary black hole merger, \citet{Bird:2016dcv} and \citet{Sasaki:2016jop} pointed out that the merger rate implied by LIGO's discovery is potentially consistent with a population of PBH accounting for all of dark matter, advancing the possibility that the two black holes involved had a primordial origin---and, indeed, that LIGO had detected dark matter.

Regardless of the relation to cosmological dark matter, the confirmation of a primordial origin for any black hole would carry great implications. Such a population might probe primordial fluctuations at scales well beyond the reach of other experiments, providing a unique window into early cosmology, and an unprecedented test of physics at extremely high energies. The primordial-origin scenario for the black holes observed at LIGO has thus been discussed heavily in the literature. Several authors have proposed that stellar- and primordial-origin models might be distinguished statistically in the coming years by the distributions of binary masses, spins, and eccentricities \citep[see e.g.][]{Cholis:2016kqi,Fernandez:2019kyb,DeLuca:2020bjf,DeLuca:2020qqa,Dolgov:2020xzo}. However, an extensive literature shows that the binaries observed to date are compatible with a stellar origin \citep{Belczynski:2009xy,Belczynski:2017gds}, and efforts to attribute any future discrepancies to a primordial origin will be complicated by uncertainties in stellar evolution models \citep[see e.g.][]{Gerosa:2017kvu,Farr:2017gtv}. Thus, even in the most optimistic case, it will be difficult to positively establish a non-stellar origin for the LIGO black holes, especially if such a history applies only to a subcomponent of the merging population.

But there is one clean signal that could clearly indicate the primordial origin of a specific black hole: a low mass. Stellar evolution models predict that black holes form only when a star's mass is sufficient for the gravitational force to overcome degeneracy pressure. Thus, black holes with a stellar origin must have a mass no lower than the Chandrasekhar limit of $\SI{1.4}{M_\odot}$ \citep{Chandrasekhar:1931ih}. A black hole with a mass below $\sim\SI{1}{M_\odot}$ must have a non-stellar origin, and the detection of even one such object would be a clear smoking gun of new physics, as was already pointed out by \cite{Chapline:1975ojl}. In principle, LIGO may be sensitive to mergers of black holes well below this scale, so LIGO and other gravitational wave observatories are uniquely capable of directly establishing the existence of PBH.

Indeed, some gravitational wave detections have already come tantalizingly close to furnishing such a discovery. The latest hint comes from the recently announced GW190814 \citep{Abbott:2020khf}, apparently involving a compact object at $\SI{2.6}{M_\odot}$, in what was expected to be a mass gap in the population of neutron stars and stellar-origin black holes \citep{Bailyn:1997xt,Ozel:2010su,Ozel:2012ax,Farr:2010tu}. Additionally, the nature of the compact objects involved in GW170817 \citep{TheLIGOScientific:2017qsa} is uncertain: while the identification of an associated kilonova \citep{Soares-Santos:2017lru} strongly indicates that one of the merging objects was a neutron star, the second compact object might also be a light $\mathcal O(\SI{1}{M_\odot})$ black hole, with likelihood as large as 40 per cent \citep[see e.g.][]{Coughlin:2019kqf}.

Given the potential for discovery, the LIGO Collaboration has conducted initial searches for mergers of light PBH \citep{Abbott:2005pf,Authors:2019qbw}, with null results thus far. But interpreting these null results as constraints on the PBH population requires a model to connect the abundance and mass distribution of PBH to the rate of observed mergers. Theoretical uncertainties in the merger rate complicate such an analysis, with notable recent progress by \cite{Vaskonen:2019jpv}. Even so, most previous work has assumed that the PBH mass function is monochromatic, i.e., that all PBH have a single mass. This greatly simplifies the problem, but is likely unrealistic: in most formation models, PBH have an extended mass distribution with a lognormal or power-law shape \citep{Carr:2016drx}. In some scenarios, the mass distribution can even be multimodal \citep{Carr:2018poi}.

A bias-free interpretation of LIGO results requires that we allow for some freedom in the shape of the mass function. This motivates the approach taken by \cite{Chen:2019irf}, who analyze prospects for the detection of light black holes under the assumption that the mergers observed thus far have a primordial origin. To further complicate matters, the mass distribution is subject to various observational constraints across the mass spectrum, which impose additional restrictions on the space of mass functions. The uncertainty in the merger rate arising from the shape of the mass function means that it is difficult to describe prospects for either constraints on or discovery of a PBH population at gravitational wave observatories in a model-independent fashion.

In this work, we use numerical methods to translate null searches at gravitational wave observatories into constraints on the properties of the PBH population and discovery prospects for light black holes. In particular, we show that if only a small fraction of the PBH population lies in the mass window of interest, then freedom in the mass function translates to a significant gap between the \emph{constraint} potential and the \emph{discovery} potential, corresponding to the most pessimistic and optimistic calculations of the merger rate, respectively. We show that LIGO can establish the existence of PBH even if the abundance of such objects in the mass range of interest is far below the level of the prospective constraint. Our results provide the first model-independent gravitational wave constraints on the light black hole population, and show that there is considerable opportunity for their discovery at LIGO.

This paper is organized as follows. In \cref{sec:merger-rate}, we review the calculation of the merger rate and establish analytical expectations for the shapes of mass functions that maximize and minimize the merger rate. In \cref{sec:optimization}, we introduce our numerical procedure and detail the inclusion of other observational constraints. We present our numerical results in \cref{sec:results}, and we discuss the implications and conclude in \cref{sec:discussion}.

Notationally, we will say that a black hole is `light' if it has a mass below $\SI{1}{M_\odot}$, and we will say that a black hole is `detectable' if it has a mass large enough to be observable by LIGO, a condition we will detail in subsequent sections. We will refer to light black holes as `primordial', although, as we have mentioned, there are other new physics scenarios that may also result in the formation of observable black holes without stellar progenitors. Our interest is in light black holes that are `detectable and primordial', which we will abbreviate as `\DPBH'. We will say that such black holes lie in the `\DP' mass range.

\section{The merger rate of \DPBH}
\label{sec:merger-rate}
To establish the most optimistic discovery prospects, and the most pessimistic constraint potential, it is necessary to consider, respectively, the maximum and minimum merger rates that can be produced with a fixed abundance of PBH. We will perform this optimization numerically in the following sections, but first, we discuss the calculation of the merger rate and explore a few benchmark cases analytically.

The merger rate of PBH has been studied by many authors \citep{Bird:2016dcv,Mandic:2016lcn,Sasaki:2016jop,Clesse:2016ajp,Raidal:2017mfl,Chen:2018czv,Vaskonen:2019jpv}, and while predictions of the rate are still subject to some uncertainties, the theoretical formalism has improved considerably in recent years. In particular, \citet{Raidal:2017mfl} and \citet{Chen:2018czv} have studied the merger rate for extended mass functions, and established predictions for the merger rate as a function of the component masses. The formation of merging primordial black hole binaries is quite different from the stellar case, so we will shortly review the derivation of the merger rate and the attendant physics.

Throughout the following sections, we will denote the mass function by $\psi$. Denoting the PBH number density for masses up to $m$ by $n(m)$, the mass function is defined by $\psi(m)\propto m\dd n/\du m$ with the normalization condition
\begin{equation}
    \int\du m\,\psi(m) = \frac{\OmegaPBH}{\OmegaDM} \equiv \fpbh.
\end{equation}

\subsection{The detectable mass range}
A key component of estimating the \DPBH~merger rate is defining exactly what is meant by detectability. In previous studies of \DPBH~mergers, a threshold is generally set at a mass of order $\sim\SI{0.1}{M_\odot}$, and mergers of black holes below the threshold mass are assumed to be undetectable. We must do the same in this work, for reasons we will explain shortly. For the moment, note that this is a reasonable approximation, especially because gravitational wave detectors trigger on the basis of a bank of template waveforms. Thus, even if LIGO is potentially sensitive to mergers of lighter objects, a detection will not be made if no matching template has been computed. In typical operation, LIGO uses no templates with combined binary masses below $\SI{2}{M_\odot}$ \citep{Magee:2018opb}, and even past searches for light black hole mergers have used a minimum template mass of $\SI{0.4}{M_\odot}$ \citep{Abbott:2005pf,Authors:2019qbw}.

Neglecting templates, LIGO is potentially sensitive to mergers of very light black holes, with one important caveat: the lighter the binary, the closer it must be in order for the merger to be detectable. Thus, LIGO probes a different effective volume $V_{\mathrm{eff}}(m_1,m_2)$ for each pair of component masses $(m_1,m_2)$. Given a particular mass function $\psi$, the total \DPBH~merger rate $R_{\mathrm{\DP}}$ must then be written in the form
\begin{equation}
    \label{eq:integrated-rate}
    R_{\mathrm{\DP}}(\psi) = \int_{\mathrm{\DP}^2}\du m_1\dd m_2\,
        \mathcal R(m_1,m_2) V_{\mathrm{eff}}(m_1,m_2),
\end{equation}
where $\mathcal R(m_1,m_2)$ is the differential merger rate per unit volume for binaries with component masses $(m_1,m_2)$, and $\int_{\mathrm{\DP}^2}$ denotes an integral only over pairs of masses in the \DP~regime. We make the simplifying approximation that the sensitive volume depends only on the chirp mass of the binary, $\mchirp$, and not on the individual component masses, so that $V_{\mathrm{eff}}(m_1,m_2)=V_{\mathrm{eff}}(\mchirp(m_1,m_2))$. This approximation has been explicitly validated by \cite{Authors:2019qbw}. Note that we neglect any impact of binary spin on detectability.

We determine the sensitive volume $V_{\mathrm{eff}}(m_1,m_2)$ for the merger of a given binary using the maximum sensitive distance for the scenario considered in \citet{Magee:2018opb}. This sensitivity is already achievable with the Advanced LIGO instrument, but does involve optimistic assumptions about the template bank used to identify merger events. The number $N$ of templates required for a given search depends strongly on the minimum mass $m_{\mathrm{min}}$ and starting frequency $f_{\mathrm{min}}$ included in the template bank, scaling as $N\propto (m_{\mathrm{min}}f_{\mathrm{min}})^{-8/3}$. We follow the optimistic benchmark of \citet{Magee:2018opb}, choosing $f_{\mathrm{min}}=\SI{10}{\hertz}$ and $f_{\mathrm{max}}=\SI{2048}{\hertz}$, and we likewise reduce the maximum sensitive distance by a factor of 2.26 to account for variations in the location and orientation of the binary. (See fig. 2 and eq. (12) of that reference.) As noted in that work, there are significant computational costs to producing an appropriate template bank for detection of these very low-mass binaries at the greatest sensitive distances. LIGO searches completed to date use slightly less generous template banks, and in particular are completely insensitive to black holes below \SI{0.2}{M_\odot} \citep{Authors:2019qbw}. Thus, our results apply directly under the assumption that LIGO carries out a search with these parameters.

We still need to define the domain of the integral in \cref{eq:integrated-rate}. To meaningfully probe the abundance of light PBH, we will ultimately be interested in speaking of the abundance in a narrow mass range, neither too massive to be clearly primordial, nor too light to be typically detectable, but just right \citep[see e.g.][]{Hassall:1904}. To that end, we will define two thresholds $\mdpmin$ and $\mdpmax$. For single masses, we will say $m\in\mathrm\DP$ if $\mdpmin\leq m\leq \mdpmax$. For pairs of masses, we will say that $(m_1,m_2)\in\mathrm{\DP}^2$ if $\mdpmin\leq\min\{m_1,m_2\}\leq\mdpmax$, i.e., if
\begin{enumerate}
    \item \emph{both} $m_1$ and $m_2$ are above $\mdpmin$, and
    \item \emph{at least one} of $m_1$ and $m_2$ is below $\mdpmax$.
\end{enumerate}
We will fix $\mdpmax=\SI{1}{M_\odot}$ and $\mdpmin=\SI{0.1}{M_\odot}$ throughout our analysis. We have investigated the consequences of choosing $\mdpmin=\SI{0.01}{M_\odot}$, and found that there is very little impact on the qualitative outcomes of our analysis: while choosing a lower threshold threshold extends the opportunity for discovery if PBH only exist at lower masses, extant gravitational wave detectors are relatively poorly suited to probe such a population.

In order to meaningfully discuss constraints on the \DPBH~population, we define the \DP~ratio by
\begin{equation}
    \rdp = \frac{1}{\fpbh}\int_{\mdpmin}^{\mdpmax}\du m\,\psi(m).
\end{equation}
This is the mass fraction of PBH with masses between $\mdpmin$ and $\mdpmax$. Note the use of $r$ (ratio) rather than $f$ to avoid confusion with $\Omega_\mathrm{\DP}/\OmegaDM$, as with $\fpbh$. We instead define $\fdp\equiv\rdp\fpbh$.

Ultimately, we will evaluate maximum and minimum merger rates as a function of both $\fpbh$ and $\rdp$ simultaneously. This is a convenient parametrization for discussing constraints on the mass function, since despite the very simple form of the two parameters, they encode key information about the abundance of PBH in general and the abundance of light black holes in particular. This is also one of the reasons for imposing a strict cut-off at low masses: one might contend that lighter black holes, with masses below our $\mdpmin$, are also detectable, albeit in a smaller volume. This may indeed be the case, but including such mergers would make the parametrization discussed here difficult to interpret in relation to the merger rate: black holes just below $\mdpmin$ would contribute to the \DP~merger rate, but not to $\rdp$.

We reiterate that LIGO is not equipped with templates for our entire \DP~window during its regular operation, and a search with black hole masses below \SI{0.2}{M_\odot} has not been conducted to date. Moreover, previous searches have targeted mergers between two light black holes, with templates only below \SI{4}{M_\odot} in total binary mass. Thus, the constraints we draw in this work are prospective, assuming that an extended search is conducted on archival or future data. As we will show, such searches are well motivated both for pairs of light black holes and for mergers of light black holes with heavy partners. There is ample opportunity to discover PBH even at abundances that cannot be fully constrained.

\subsection{Estimating the merger rate}
We now review the derivation of the merger rate in \citet{Raidal:2017mfl} and \citet{Chen:2018czv}. First, we note that PBH binaries can form in two epochs: in the early Universe, during radiation domination, and in the late Universe, where close approaches can produce enough gravitational radiation to bind two black holes. The latter contribution is generally small, since typical relative velocities are large, meaning that the energy loss to gravitational radiation must be quite significant. There is a possible exception to this rule if the density contrast in the late Universe is exceptionally large, $\delta_0\gtrsim 10^{10}$, but this is much larger than most estimates, so we neglect that possibility. Thus, we consider only binaries formed in the early Universe. Note that our calculation may not accurately describe new physics scenarios in which light black holes themselves form in the late Universe.

We first review the merger rate as estimated by \citet{Raidal:2017mfl}. Consider a PBH pair with masses $m_1$ and $m_2$. First, in order for the pair to decouple from the Hubble flow and have interactions dominated by their mutual gravitation, the average mass of the black holes should exceed the background mass in the volume between them, i.e., we require $\frac12(m_1+m_2)>\frac{4\pi}{3}\rho_{\mathrm{bg}}r^3$. Translating this into a condition on the separation of the two black holes, one finds that the comoving distance between them must fall below the scale
\begin{equation}
    \tilde x(m_1,m_2)^3 =
        \frac{3}{4\pi}\frac{m_1+m_2}{a_{\mathrm{eq}}^3\rhoeq},
\end{equation}
where $a_{\mathrm{eq}}$ and $\rhoeq$ are the scale factor and the density at matter--radiation equality. A binary with comoving separation $x<\tilde x$ thus decouples from the Hubble flow when the scale factor is
\begin{equation}
    a_{\mathrm{dc}} = a_{\mathrm{eq}}\left(\frac{x}{\tilde x}\right)^3.
\end{equation}
After this point, the black holes' gravity dominates the evolution of the system. Barring a close approach, gravitational interactions between these two black holes and some third body of comparable mass are necessary to move the pair into a bound configuration. Thus, we suppose that there is third PBH with mass $m_3$ at a comoving distance $y$ from the first two. In this scenario, we form a binary with semimajor and semiminor axes given by 
\begin{equation}
    r_a = \alpha x a_{\mathrm{dc}},
    \qquad
    r_b = \beta \frac{2m_3}{m_1+m_2}\left(\frac xy\right)^3r_a,
\end{equation}
for two $\mathcal O(1)$ constants $\alpha$ and $\beta$. We take $\alpha=\beta=1$ for the remainder of this discussion. Assuming that there is no mechanism for hardening the binary apart from gravitational radiation, the coalescence time can then be estimated as 
\begin{equation}
    \tilde\tau(m_1,m_2,m_3)\left(\frac{x}{\tilde x(m_1,m_2)}\right)^{37}\left(\frac{y}{\tilde x(m_1,m_2)}\right)^{-21},
\end{equation}
where $\tilde\tau$ is the maximal coalescence time, given by 
\begin{equation}
    \tilde\tau(m_1,m_2,m_3) = \frac{348}{85}\frac{
        \alpha^4\beta^7 a_{\mathrm{eq}}^4 m_3^7 \tilde x(m_1,m_2)^4
    }{
        G^3m_1m_2(m_1+m_2)^8
    }.
\end{equation}
Here we have established the coalescence time for a single binary assuming a set of masses and initial separations. Distributions of these parameters can be derived from the mass function, leading to a distribution of coalescence times as a function of the component masses. Differentiating this distribution leads to the merger rate at the present time. For brevity, we define $\tilde m(\psi)=1/\int\du m\,\psi(m)/m$. The number density of PBH at the mass of interest is accounted for through the factor $\tilde N(\psi;\;m_1,m_2) = \delta_{\mathrm{dc}}\Omega_{\mathrm{DM,eq}}(m_1+m_2)/\tilde m(\psi)$, where $\delta_{\mathrm{dc}}$ is the density contrast at the time of decoupling. We then define
\begin{multline}
    \mathcal G(\psi;\; m_1,m_2,m_3)=\Gamma\left(
        \frac{58}{37},\;
        \frac{\tilde N(\psi;\;m_1,m_2)\,t^{3/16}}{\tilde\tau(m_1,m_2,m_3)^{3/16}}
    \right) \\
    -\Gamma\left(
        \frac{58}{37},\;
        \frac{\tilde N(\psi;\;m_1,m_2)\,t^{-1/7}}{\tilde\tau(m_1,m_2,m_3)^{-1/7}}
    \right),
\end{multline}
and the present-day differential merger rate between black holes with masses $m_1$ and $m_2$ is given by
\begin{multline}
    \mathcal R(m_1, m_2) =
    \frac{
        9\bar m(\psi)^3
        \tilde N(\psi;\;m_1,m_2)^{\frac{53}{37}}
    }{
        296\pi \delta_{\mathrm{dc}}\tilde x(m_1,m_2)^{3}t^{34/37}
    }
    \\\times
    \frac{\psi(m_1)\psi(m_2)}{m_1m_2}\int\du m_3\,
        \frac{\mathcal G(\psi;\;m_1,m_2,m_3)}{\tilde\tau(m_1,m_2,m_3)^{3/37}}
        \frac{\psi(m_3)}{m_3}.
\end{multline}
Notably, this estimate of the merger rate considers only the tidal torque due to one additional PBH external to the binary. This may present a problem when dealing with mass functions that span many decades, for which lighter black holes have relatively high number densities. \citet{Chen:2018czv} follow a similar line of argument, but the authors estimate the torque by integrating over the entire PBH population. It might be expected that this form of the merger rate is more reliable for extremely broad or multimodal mass functions, which we may well encounter in the course of our analysis. Thus, we use their merger rate in the course of our calculation, and we now briefly summarize their result. We define
\begin{equation}
    \mu = \frac{
        2m_1m_2\left(\psi(m_1)+\psi(m_2)\right)
    }{
        (m_1+m_2)\bigl(m_1\psi(m_1)+m_2\psi(m_2)\bigr)
    }
\end{equation}
and $n_T=\rho_c\Omega_{\mathrm{DM,eq}}\int\du m\,\psi(m)/m$, where the lower limit of integration is $\min\{m_1,m_2\}$. We additionally take $\langle x\rangle$ to be the average separation between black holes of mass $m_1$ and $m_2$, and define $\gamma_X$ by
\begin{multline}
    \gamma_X = \left(
        \frac{85}{3}\frac{
            tm_1m_2(m_1+m_2)\bigl(\psi(m_1)+\psi(m_2)\bigr)^4
        }{
            10^{-4}\left(
                \frac{3}{8\pi}\frac{
                    m_1+m_2
                }{
                    \rhoeq(\psi(m_1)+\psi(m_2)
                }
            \right)^{4/3}X^{16/3}
        }
    \right)^{1/7}
    \\\times
    \frac{
        2(\psi(m_1)+\psi(m_2))\Omega_{\mathrm{M}}
    }{
        \Omega_{\mathrm{DM}}X
    }.
\end{multline}
Then the probability distribution for the coalescence time is given by
\begin{equation}
    \label{eq:dPdt}
    \frac{\du P}{\du t} = \frac{1}{7\mu t}
    \int\du X\,\exp\left(
        -\frac{X}{\mu}\frac{4\pi}{3}\langle x\rangle^3n_T
    \right)\frac{\gamma_X^2}{\left(1+\gamma_X^2\right)^{3/2}},
\end{equation}
and the present-day merger rate per unit volume is given differentially in the component masses by
\begin{equation}
    \label{eq:merger-rate}
    \mathcal R\left(m_1,m_2\right)=\rho_c\Omega_{\mathrm{M}}\min\left(
        \frac{\psi(m_1)}{m_1},\;\frac{\psi(m_2)}{m_2}
    \right)\frac{\du P}{\du t}
    .
\end{equation}

While we use this form of the merger rate in our subsequent analysis, it is not the only such calculation to take into account the torques from the entire population of black holes. The calculation of \citet{Raidal:2017mfl} was updated and extended by \citet{Raidal:2018bbj} to include this effect via a suppression factor multiplying the rate. The suppression factor $S$ has the form
\begin{align}
    &S = \frac{e^{-\bar N}}{\Gamma(21/37)}\int\du v\,v^{-\frac{16}{37}}
        \exp\left(
            - \mathcal F(\psi,v)
            - \frac{3\sigma_{\mathrm M}^2v^2}{10\fpbh^2}
        \right)
    ,\\
    &\mathcal F(\psi,v) = \bar N\langle m\rangle\int\frac{\psi(m)\dd m}{m}
        \leftidx{_1}{F}{_2}\left(
            -\tfrac12;\tfrac34,\tfrac54;-\left(
                \tfrac{3mv}{4\langle m\rangle\bar N}
            \right)^2
        \right)
    ,
\end{align}
where $\langle m\rangle$ is the average PBH mass, $\leftidx{_1}{F}{_2}$ is the generalized hypergeometric function, $\bar N$ counts the number of PBHs in the vicinity of a given binary, and $\sigma_{\mathrm M}=(\Omega_{\mathrm{M}}/\Omega_{\mathrm{DM}})^2\langle\delta_{\mathrm{M}}^2\rangle$ for $\delta_{\mathrm{M}}$ the matter density perturbation. Note that in this merger rate calculation, the suppression effect factorizes away from the dynamics that determine the binary formation rate, so the suppression factor can really be evaluated separately from the unsuppressed rate. This is not the case in the calculation of \citet{Chen:2018czv}: while there is no explicit suppression factor, a comparable suppression enters the rate itself via the exponential factor in \cref{eq:dPdt}. However, the calculation does not include a suppression of the merger rate from binary disruption in later close encounters.

We note that predicting the PBH merger rate from first principles is extremely challenging, and it is likely that these estimates will be refined in the coming years. In particular, \citet{Jedamzik:2020ypm} recently showed that the inclusion of all three-body encounters in PBH clusters can dramatically reduce the merger rate in the late Universe. We will return to this possibility in \cref{sec:discussion}, and we will also assess the robustness of our results to differences between the calculations of \citet{Chen:2018czv} and \citet{Raidal:2018bbj}.

\subsection{Analytical behavior of the merger rate}
\label{sec:analytical}
To establish constraint and discovery prospects, we will need to minimize and maximize the merger rate over the possible mass function shapes with some characteristic abundances held fixed. In particular, we will optimize the merger rate with $\rdp$ and $\fpbh$ held constant. For fixed $\fpbh$, if the maximum merger rate falls below the LIGO sensitivity for a given value of $\rdp$, this means that values of $\rdp$ this low cannot be probed by LIGO, regardless of the form of the mass function. Alternately, if the minimum merger rate is detectable by LIGO, then this and higher values of $\rdp$ can be ruled out by LIGO.

In general, the merger rate must be maximized or minimized numerically. However, to understand the dependence of the merger rate on the shape of the mass function, it is useful to consider a few simple benchmark cases in the absence of any observational constraints. For the moment, we neglect the mass dependence of the detector's sensitive volume.

\begin{figure}\centering
    \includegraphics[width=\columnwidth]{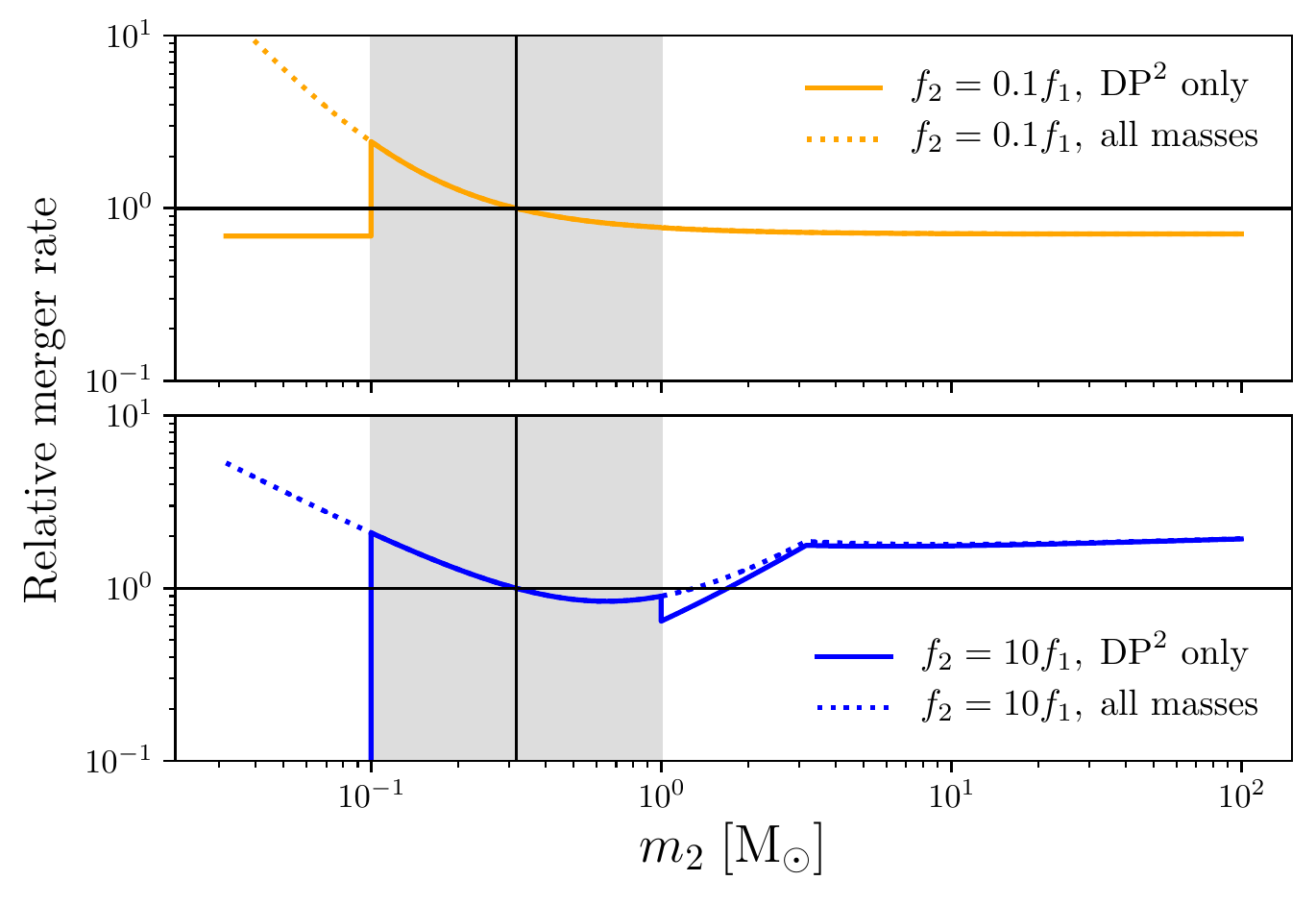}
    \caption{Merger rate for a dichromatic mass function, $\psi(m) = f_1\delta(m-m_1) + f_2\delta(m-m_2)$, relative to the monochromatic mass function $(f_1+f_2)\delta(m-m_1)$. We fix $m_1=10^{-1/2}\SI{}{M_\odot}$, indicated by the black vertical line. This lies in the middle of the \DP~window, indicated by the shaded region. The dashed curves show the merger rate for pairs of all masses, while the solid curves include only mergers in $\mathrm{\DP}^2$. The blue curve shows the case $f_2=10f_1$, i.e., where mergers of black holes of mass $m_2$ naively dominate. The orange curve shows the case $f_2=0.1f_1$, where the reverse is true. Depending on the relative amplitudes and positions of the two peaks, separating them can either enhance or suppress the merger rate (see text).}
    \label{fig:dichromatic}
\end{figure}

\begin{figure}\centering
    \includegraphics[width=\columnwidth]{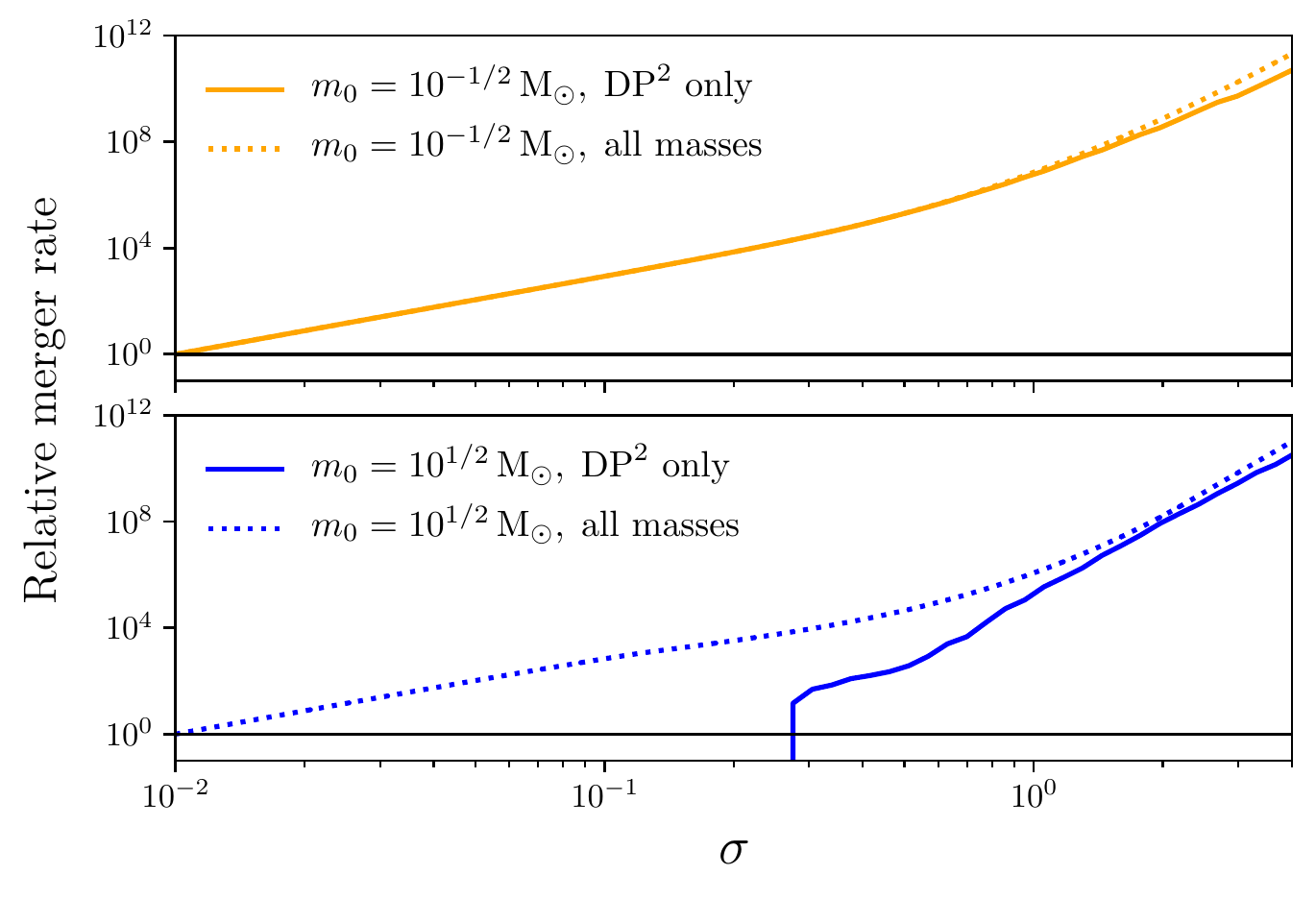}
    \caption{Relative merger rate as $\sigma$ is increased for a lognormal mass function with two different central masses. The dashed lines include all mergers, while the solid lines include only $\mathrm{\DP}^2$ mergers. The curves are normalized relative to the all-inclusive merger rate at the lowest value of $\sigma$, i.e., the dashed lines are fixed to 1 at the left edge of the figure. For much of the range of $\sigma$ shown here, a large fraction of mergers lie in $\mathrm{\DP}^2$, so the solid and dashed lines lie very close. They begin to diverge at large $\sigma$ in both subplots, since much of the mass lies outside the \DP~window in this case. For the blue curve, the central mass lies outside the \DP~window, so the \DP~merger rate vanishes at small $\sigma$. In the limit $\sigma\to0$, the lognormal mass function reduces to the monochromatic case.}
    \label{fig:lognormal}
\end{figure}

First, consider a monochromatic mass function, $\psi(m)=f_1\delta(m-m_1)$. Formally, the quantities entering \cref{eq:merger-rate} are not independently well defined in this case, but we can take a mass function of the form
\begin{equation}
    \label{eq:delta-before-limit}
    \psi_1(m) = f_1\Delta^{-1}\Theta(m-m_1)\Theta(m_1+\Delta-m)
\end{equation}
and work in the limit $\Delta\to 0$. In this case, the total \DP~rate is simply the overall rate, as long as $m_1$ lies within the \DP~window. If $m_1$ is sufficiently small, the integrand of \cref{eq:dPdt} is dominated by values of $X$ where the exponential is very nearly 1. As pointed out by \cite{Chen:2018czv}, the integral can then be evaluated approximately, which gives a rate $\mathcal R\propto m_1^{-32/37}$. A similar result can be derived for large $m_1$ by splitting the $X$ integral into two regimes, one in which the constant term in the denominator is dominant, and one in which it is subdominant. The integral can be evaluated analytically in each of the two regimes, and it can then be shown that $\mathcal R\propto m_1^{-26/21}$ at large $m_1$.

In particular, for a monochromatic mass function, decreasing $m_1$ increases the merger rate. Physically, this is simply because decreasing $m_1$ while holding $f_1$ constant increases the number density of black holes. In the absence of observational constraints, we therefore expect that the merger rate will be maximized when the mass function is peaked near the bottom of the \DP~window, and minimized when it is peaked near the top.

This is the simplest way in which the mass function can influence the merger rate. However, monochromatic mass functions are tightly constrained by observational bounds, so it is useful to understand the behavior of the merger rate for mass functions with non-negligible width. We first consider the simplest extension of the previous case: a bimodal mass function constructed as the sum of two monochromatic mass functions. We define
\begin{equation}
    \psi_2(m) = f_1\delta(m-m_1) + f_2\delta(m-m_2),
\end{equation}
where the Dirac delta is understood to be defined as in \cref{eq:delta-before-limit}.

For such a `dichromatic' mass function, there are three contributions to the merger rate, corresponding to mergers of black holes with masses $\{m_1, m_1\}$, $\{m_2, m_2\}$, and $\{m_1,m_2\}$. This gives rise to complicated behavior as the peaks are separated. Two benchmark cases are shown in \cref{fig:dichromatic}, with one peak fixed in the middle of the \DP~range and the other varying freely. In each panel, the merger rate is enhanced if the second peak is positioned at a low mass within the \DP~window, due to the enhanced number density. The \DP~merger rate (the solid line) drops sharply as the mass of the second peak falls below the \DP~window, while the all-inclusive merger rate (the dotted line) continues to increase.

Notice that as the second peak rises above the \DP~window, the drop in the \DP~merger rate is much less significant. This is because the presence of these more massive black holes still affects the \DP~merger rate in two ways: first, more massive black holes can still participate in the formation of light PBH binaries, and secondly, mergers of binaries with masses $\{m_1,m_2\}$ themselves contribute to the \DP~merger rate. These effects lead to non-trivial behavior of the dichromatic merger rate as a function of the two masses. For our purposes, we note that separating peaks in a dichromatic mass function can either increase or decrease the merger rate.

Finally, we consider a lognormal mass function, which is unimodal, but has a non-vanishing width. The lognormal mass function has the form
\begin{equation}
    \psi_L(m) = \frac{\fpbh}{\sqrt{2\pi}\sigma m}\exp\left(
        -\frac12\left[\frac{\log\left(m/m_0\right)}{\sigma}\right]^2
    \right),
\end{equation}
where $m_0$ corresponds to the central mass and $\sigma$ is the width of the distribution. Holding $\sigma$ fixed, the merger rate is increased by reducing $m_0$, as in the monochromatic case. If $m_0$ is fixed, and $\sigma$ is varied, then the merger rate is enhanced by increasing $\sigma$, i.e., broadening a sharp distribution locally increases the merger rate. This behavior is shown in \cref{fig:lognormal}.

These benchmark scenarios indicate that a fairly broad mass function favouring lower masses will generally produce a higher merger rate, but in general, observational constraints will impose severe restrictions on the allowed shape of the mass function. Thus, the mass function that minimizes or maximizes the merger rate might indeed be a complicated multimodal function. In particular, the analysis of \cite{Lehmann:2018ejc} demonstrates that the maximum value of $\fpbh$ consistent with observational constraints is attained by a multimodal mass function, corresponding to a superposition of monochromatic mass functions. Thus, it would not be surprising to find a similar behavior for mass functions that maximize the merger rate, particularly if $\fpbh$ is fixed at a value where observational bounds strongly constrain the mass function. The benchmark scenarios in this section suggest that the merger rate will be minimized with sharply peaked and potentially multimodal mass functions.

To go further and to incorporate observational constraints will require numerical methods, which we take up in the following sections. Again, note that the discussion above has not accounted for any characteristics of the detector. In particular, we have neglected the mass dependence in the effective volume that an instrument such as LIGO can probe. Mergers of more massive black holes are observable in a larger effective volume, and this enhances the effective merger rate at higher PBH masses, competing with the enhancement in number density at lower masses. We will include this effect in our numerical treatment.

\section{Constraints and optimization}
\label{sec:optimization}
In this section, we detail the numerical procedure that we use to optimize the merger rate. First, we discuss the constraints that we impose on the black hole mass function.

Since the allowed values of the merger rate depend on the allowed forms of the PBH mass function, observational constraints that restrict the form of the mass function correspondingly restrict the merger rate. Thus, the minimum or maximum merger rate is dependent not only on $\fpbh$ and $\rdp$, but also on the chosen set of observational constraints. The full set of observational constraints we use in this work is shown in \cref{fig:free-optima,fig:constrained-optima}, with descriptions in the captions. We demonstrate the behavior of the maximum and minimum merger rates both with and without the constraints imposed on the mass function by these observational bounds.

Note that there are many other observables that may place constraints on the PBH population, such as supernova lensing \citep{Zumalacarregui:2017qqd}, dynamical effects \citep{Monroy-Rodriguez:2014ula,Brandt:2016aco,Koushiappas:2017chw}, and destruction of white dwarf stars \citep{Graham:2015apa} and neutron stars \citep{Capela:2013yf}. (For reviews see \cite{Carr:2016drx,Carr:2020gox,Carr:2020xqk}.) These constraints are subject to additional uncertainties, and including them does not change our qualitative conclusions. The qualitatively important features are the relative strength of the constraints at masses above and below the \DP~window, and the fact that there is a gap in the constraints at low masses. The latter allows for large values of $\fpbh$ when $\rdp$ is small. This gap has attracted considerable attention since lensing constraints at low masses were shown to be ineffective \citep{Inomata:2017vxo,Katz:2018zrn,Montero-Camacho:2019jte,Smyth:2019whb,Sugiyama:2019dgt}, and it is possible that new constraints developed in this region will influence our results.

We introduce one important observable beyond the constraints plotted in \cref{fig:free-optima,fig:constrained-optima}: the stochastic gravitational wave background \citep[SGWB;][]{Abbott:2009ws,TheLIGOScientific:2016dpb}. A population of black holes merging over cosmic time produces an accumulated background of gravitational radiation that can be detected by LIGO. Since the SGWB depends in detail on the shape of the mass function, it must be treated differently from the other constraints. However, it is essential that we include this constraint, since it has been shown that merging \DPBH~in particular can make a significant contribution to the SGWB \citep[see e.g.][]{Wang:2019kaf}. Further, when we maximize the merger rate, we also maximize the contribution to the SGWB, so our optimal mass functions might run afoul of SGWB constraints at PBH abundances well below those excluded in other analyses.

\subsection{Applying constraints to the mass function}
In order to translate gravitational wave observables to discovery prospects and constraints on the population of \DP~black holes, we must alternately minimize and maximize the merger rate subject to particular constraints. This is similar to the problem of maximizing the overall abundance of black holes subject to observational constraints, as discussed by \citet{Lehmann:2018ejc}. In that reference, the general form of the optimal mass function is derived analytically, and it is shown that the exact global optimum can be found semi-analytically with arbitrary precision. Since we will use some of the same methods and terminology, we briefly review this result. However, as we will explain, this formalism cannot be adapted to optimize the merger rate semi-analytically.

We treat observational constraints on the black hole population following \cite{Carr:2017jsz}. In general, observational constraints on the black hole population are derived from some measured quantity $A_{\mathrm{obs}}$. The value of $A_{\mathrm{obs}}$ is predicted to be $A_0$ in the absence of any PBH, whereas in the presence of PBH, one has $A_{\mathrm{obs}}=A_0+A_1$. For most observables, black holes at different mass scales contribute independently, so we can write
\begin{equation}
    \label{eq:constraint-form}
    A_1 = \int\du m\,\psi(m)\,K_1(m)
\end{equation}
for some kernel $K_1(m)$. Provided that the constraining observable has this form, the constraint condition can be written in the form $\mathcal C(\psi)\leq1$, where $C(\psi)$ is the functional
\begin{equation}
    \mathcal C(\psi)\equiv\int\du m\,\frac{\psi(m)}{\fmaxmono(m)}.
\end{equation}
Here $\fmaxmono(m)$ is the maximum allowed fraction of dark matter in the form of PBH assuming a monochromatic mass function at mass $m$. For the case of $N$ independent constraints $f_{\mathrm{max},j}(m)$, corresponding to a vector $\mathcal C_j(\psi)$, this generalizes to 
\begin{equation}
    \label{eq:constraint-functional}
    \bigl\|\cvec(\psi)\bigr\|^2\equiv \sum_{j=1}^N\left(\int\du m\,\frac{\psi(m)}{f_{\mathrm{max},j}(m)}\right)^2 \leq 1.
\end{equation}
Note, in particular, that $\psi(m)>\fmaxmono(m)$ is perfectly admissible for a subset of masses---i.e., the mass function can cross through the curves on constraint plots---as long as the condition above is still met. This is simply because constraint curves, as typically drawn, are only applicable to monochromatic mass functions.

Since the total density in PBH scales linearly with the normalization of the mass function, any mass function can be normalized to saturate observational constraints, yielding the \emph{normalized mass}, $\mathcal M(\psi) = \|\cvec(\psi)\|^{-1}\int\du m\,\psi(m).$ Finding the mass function that maximizes the PBH density subject to observational constraints is thus equivalent to maximizing the functional $\mathcal M$, which can be done semi-analytically.

One might hope that a similar method might apply to the optimization of the merger rate. But even if the merger rate functional were as simple as the normalized mass, it would still not be possible to apply the preceding formalism. As we have discussed, it is essential to consider constraints from non-detection of a SGWB signal, but this constraint cannot be cast in the form of \cref{eq:constraint-functional}. We now briefly review the nature and calculation of the SGWB constraint.

A population of PBH produces a SGWB from mergers at higher redshifts \citep{Regimbau:2011rp,Rosado:2011kv,Zhu:2011bd,Wang:2016ana,Wang:2019kaf}. While such backgrounds do not furnish a smoking-gun signal of a primordial origin for a particular black hole, they do constrain the PBH mass function. A differential merger rate $\mathcal R$ produces a stochastic background at frequency $\nu$ with density
\begin{multline}
    \OmegaGW(\nu) = \frac{\nu}{\rho_c}
        \int \du z\dd m_1\dd m_2\,
            \frac{\mathcal R(z;m_1,m_2)}{(1+z)H(z)}
            \\ \times
            \frac{\du E_{\mathrm{GW}}}{\du\nu_s}(\nu_s;m_1,m_2),
\end{multline}
where $\rho_c$ is the critical density and $\du E_{\mathrm{GW}}/\du\nu_s$ denotes the spectrum of the radiation emitted during a merger, with $\nu_s=(1+z)\nu$ denoting the frequency at the source. We follow the computation of the spectrum and the resulting $\OmegaGW(\nu)$ in \cite{Zhu:2011bd} and \cite{Wang:2019kaf}. LIGO is most sensitive to the SGWB at a frequency of $\nu_p\sim\SI{20}{\hertz}$, and the sensitivity is sharply peaked around $\nu_p$. Thus, we determine whether a mass function is ruled out by SGWB production by simply comparing $\OmegaGW(\nu_p)$ with LIGO constraints at that frequency, translating to the requirement that $\OmegaGW(\nu_p)\lesssim2\times10^{-9}$ \citep{Wang:2019kaf}.

Since the calculation of the SGWB is dependent on the shape of the entire mass function, this constraint cannot be expressed in the form of \cref{eq:constraint-form}. In particular, note that the strength of the constraint is not linear in the normalization of the mass function, since the merger rate itself depends on the normalization in a highly non-linear fashion. There is no simple closed-form rescaling of the mass function that will saturate SGWB constraints. Practically, this is not an issue since the optimal mass functions and the corresponding constraints on the black hole population must ultimately be derived numerically rather than analytically. In our numerical procedure, we can incorporate SGWB constraints on a nearly equal footing with other observational constraints, as we will explain in the following section.

\subsection{Numerical procedure}
\label{sec:numerics}
We now detail the numerical procedure that we use to optimize the merger rate. We minimize and maximize the merger rate using simulated annealing \citep{Kirkpatrick:1983zz}. In simulated annealing, at each step of the algorithm, a random modification to the state of the system is generated. Each modification is probabilistically accepted or rejected, and steps that decrease the cost function are preferrentially accepted---i.e., simulated annealing is a Monte Carlo Markov chain (MCMC) optimization algorithm. Simulated annealing is structurally similar to the Metropolis--Hastings algorithm \citep{Metropolis:1953am,Hastings:1970aa} for drawing samples from a distribution, but the probability of accepting a given step changes over time.

Heuristically, simulated annealing is based on an analogy to the physical process of annealing, in which a material is heated and then cooled slowly to relieve internal stresses. Heating allows the material to return to an equilibrium configuration, and since the cooling is slow, the material is likely to be in or near its equilibrium state once frozen. In simulated annealing, the system is first `heated' in the sense that random steps are accepted with a high probability. Then the temperature is slowly reduced, so that the system increasingly disfavours departure from equilibrium. This procedure locates global optima relatively efficiently: at first, while the system is `hot', the algorithm can generate a chain that explores the parameter space broadly, with little chance of being stuck at a local optimum. As the system cools, the chain is less likely to depart from a local optimum, so it tends to locate that optimum more precisely with subsequent steps.

\subsubsection{The annealing algorithm}
The simulated annealing procedure is outlined in \cref{alg:annealing}. The mass function $\psi(m)$ is binned into a set of values $\psi_i$ with bin widths $\Delta m_i$, so that $\fpbh=\sum_i^N\psi_i\,\Delta m_i$.

The number of mass bins, $N$, must be large enough to allow for sufficient flexibility in the mass function, but must not be so large as to make the calculation intractable. The computational cost of the merger rate calculation scales asymptotically as a power law in $N$, but more importantly, each additional mass bin constitutes an additional dimension for the optimization problem. Naively, since the size of a discretized search space scales exponentially with the number of dimensions, one expects a similar behavior for the number of steps to convergence of the optimization algorithm, i.e., $n_{\mathrm{steps}}\sim b^N$. If the exponential base $b$ were large, the numerical optimization we attempt here would be extremely challenging. Pragmatically, since values of the mass function in adjacent bins are highly correlated, $b$ is manageably small: in direct numerical experiments, by subdividing mass bins, we find that $b\sim 1.5$. We choose $N=21$, dividing the bins into three regions. We use 13 bins for $m<\mdpmin$, 5 bins for $\mdpmin<m<\mdpmax$, and 3 bins for $m>\mdpmax$, subdividing each region into equally sized logarithmic bins. This makes it feasible for us to generate the numerical results in this work with $\mathcal O(10^4)$ processor hours.

The probability of accepting (`jumping' to) the candidate step, $P_{\mathrm{jump}}$, is defined by
\begin{equation}
    P_{\mathrm{jump}} =
        \left[\algfunc{cost}(\psi^\prime)/\algfunc{cost}(\psi)\right]^{-1/T},
\end{equation}
where $\algfunc{cost}$ represents the functional to be minimized, and $T$ is the `temperature'. In the simplest case, $\algfunc{cost}$ is the \DP~merger rate (for minimization) or its negative (for maximization).
In our case, where the optimization problem is constrained, it is convenient to implement these constraints by adding terms to the cost function. Constraints appear in the cost function with a factor of $1/T$ so that, as the temperature drops, constraints become more important. Thus, our cost function is defined by
\begin{equation}
    \algfunc{cost}\left(\psi\right) =
        \pm R_{\mathrm{\DP}}(\psi) +
            \frac{\beta}{T}\max\{0, \mathcal P(\psi)\},
\end{equation}
where the penalty functional $\mathcal P$ is defined by
\begin{equation}
    \mathcal P(\psi) = 
        \exp\left(
            \max\left\{
                \bigl\|\cvec(\psi)\bigr\|,\; \OmegaGW/\OmegaGW^{\mathrm{max}}
            \right\}
            -1
        \right)
        -1,
\end{equation}
with $\cvec(\psi)$ defined as in \cref{eq:constraint-functional}. We choose $\beta=\SI{e3}{\per\year}$ so that even when the merger rate is at its maximum, the penalty functional still dominates the cost at the lowest temperatures we consider.

In addition, there are three components of the simulated annealing algorithm that must be implemented in a manner specific to each application: the selection of the initial point, the generation of new steps, and the cooling rate (annealing schedule).

To start new chains, we determine the initial mass function $\psi_0$ by choosing a random value in each mass bin from the log-uniform distribution on $[1,10^3]$. The resulting mass function is then rescaled to match the input values of $\rdp$ and $\fpbh$. The generation of new steps is represented by the $\textsc{neighbor}$ function, which mutates the current state of $\psi$ to obtain a candidate $\psi^\prime$. The behavior of $\textsc{neighbor}$ is specified in \cref{alg:neighbor}. Schematically, a step is generated by modifying the value of $\psi$ in a randomly selected bin $i$. The modification is drawn from a normal distribution with mean $\psi_i$ and standard deviation $\sigma\psi_i/\Delta m_i$. Appropriate sections of the resulting mass function are then rescaled to match the input $\rdp$ and $\fpbh$.

\begin{figure}
    \begin{algorithm}[H]
        \caption{Annealing procedure}
        \label{alg:annealing}
        \begin{algorithmic}[1]
            \State $k \gets 0$, $\psi\gets\psi_0$
            \While {$k < k_{\mathrm{max}}$}
                \State {$\psi^\prime \gets \Call{neighbor}{\psi}$}
                \Comment{Generate modification}
                \If {$P_{\mathrm{jump}}(\psi^\prime, T(k)) >
                \algfunc{random}((0, 1))$}
                    \State {$\psi \gets \psi^\prime$}
                    \Comment{Accept modification}
                    \State {$k=k+1$}
                \EndIf
            \EndWhile
        \end{algorithmic}
    \end{algorithm}
\end{figure}

\begin{figure}
    \begin{algorithm}[H]
        \caption{Neighbor generation}
        \label{alg:neighbor}
        \begin{algorithmic}[1]
            \Procedure{neighbor}{$\psi$}
                \State{
                    $\psi^\prime \gets \psi$
                }
                \State {
                    $i \gets
                        \algfunc{random}(\{1,2,\dotsc,N_{\mathrm{bins}}\})$
                }
                \Comment{Choose bin}
                \State{
                    $\psi^\prime_i \gets \psi_i\times
                        \algfunc{normal}(\psi_i,\,\sigma\psi_i/\Delta m_i)$
                }
                \Comment{Modify bin}
                \State{
                    $I\gets \sum_i\psi_i\Delta m_i$
                }
                \Comment{Fix $r_{\mathrm\DP}$}
                \State{
                    $I_{\mathrm\DP}\gets
                        \sum_{i\in\mathrm\DP}\psi_i\Delta m_i$
                }
                \For{$i\in\mathrm\DP$}
                    \State {
                        $\psi^\prime_i \gets
                            \psi^\prime_i\times
                            \rdp(I-I_{\mathrm\DP})/[I_{\mathrm\DP}(1-\rdp)]$
                    }
                \EndFor
                \State{
                    $I\gets \sum_i\psi_i\Delta m_i$
                }
                \Comment{Fix $\fpbh$}
                \For{$i=1,\dotsc,N_{\mathrm{bins}}$}
                    \State {
                        $\psi^\prime_i \gets \psi^\prime_i \times \fpbh / I$
                    }
                \EndFor
                \State{
                    \Return{$\psi^\prime$}
                }
            \EndProcedure
        \end{algorithmic}
    \end{algorithm}
\end{figure}

We use a modified exponential cooling schedule, with a lower limit of $T=1$. The temperature at the $k$th step is thus
\begin{equation}
    T(k) = 1 + (T_0 - 1)(1-\alpha)^k,
\end{equation}
where we set $\alpha=10^{-2}$, and $T_0$ is the initial temperature. In general, $T_0$ must be chosen empirically to allow the algorithm to explore a wide parameter space initially. We choose the initial temperature so that 80 per cent of steps from the initial position that increase the cost are accepted. Such a temperature is high enough to `melt' the system, allowing almost any step to proceed, while still being low enough that steps will be constrained as the temperature is lowered.

To verify convergence, we optimize the mass function five times, i.e., with five independent chains, at each parameter point. We evolve each chain for $10^7$ steps. Each of these chains begins with its own random mass function and with a high temperature, so convergence to the same optimum merger rate and mass function provides reassuring evidence that the algorithm is not stochastically settling into local optima. We find empirically that the merger rate typically converges across the chains within $\mathcal O(10^5)$ steps.

\subsubsection{Two-parameter optimization}
\label{sec:numerics-refinement}

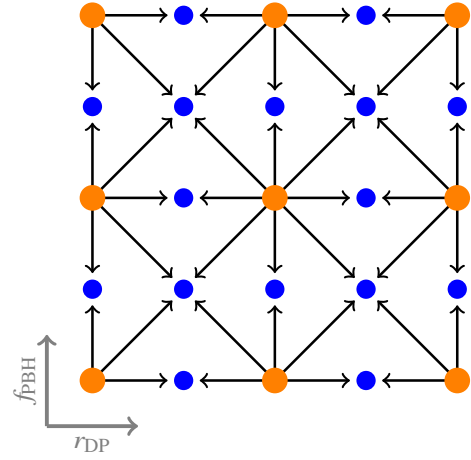
\begin{figure}\centering
    {
    \newcommand\xmin0
    \newcommand\xmax4
    \newcommand\ymin0
    \newcommand\ymax4
    \newcommand\arrow[4]{
        \draw[->,thick,shorten >=5pt] (#1, #2) -- ++(#3, #4);
    }
    \scalebox{1.2}{\begin{tikzpicture}
        \draw[->,very thick,gray] (-0.5, -0.5) --
            node[below] {$\rdp$} ++(1, 0);
        \draw[->,very thick,gray] (-0.5, -0.5) --
            node[rotate=90,left,anchor=south] {$\fpbh$} ++(0, 1);  
        \foreach \i in {\xmin,...,\xmax} {
            \foreach \j in {\ymin,...,\ymax} {
                \fill[blue] (\i,\j) circle (3pt);
            }
        }
        % Bottom left corner
        \arrow{\xmin}{\ymin}{1}{0}
        \arrow{\xmin}{\ymin}{1}{1}
        \arrow{\xmin}{\ymin}{0}{1}
        % Bottom right corner
        \arrow{\xmax}{\ymin}{-1}{0}
        \arrow{\xmax}{\ymin}{0}{1}
        \arrow{\xmax}{\ymin}{-1}{1}
        % Top right corner
        \arrow{\xmax}{\ymax}{-1}{0}
        \arrow{\xmax}{\ymax}{0}{-1}
        \arrow{\xmax}{\ymax}{-1}{-1}
        % Top left corner
        \arrow{\xmin}{\ymax}{1}{0}
        \arrow{\xmin}{\ymax}{0}{-1}
        \arrow{\xmin}{\ymax}{1}{-1}
        % Left side
        \arrow{\xmin}{{\ymax/2}}{1}{0}
        \arrow{\xmin}{{\ymax/2}}{1}{1}
        \arrow{\xmin}{{\ymax/2}}{0}{1}
        \arrow{\xmin}{{\ymax/2}}{0}{-1}
        \arrow{\xmin}{{\ymax/2}}{1}{-1}
        % Bottom side
        \arrow{{\xmax/2}}{\ymin}{1}{0}
        \arrow{{\xmax/2}}{\ymin}{-1}{0}
        \arrow{{\xmax/2}}{\ymin}{1}{1}
        \arrow{{\xmax/2}}{\ymin}{0}{1}
        \arrow{{\xmax/2}}{\ymin}{-1}{1}
        % Right side
        \arrow{\xmax}{{\ymax/2}}{0}{1}
        \arrow{\xmax}{{\ymax/2}}{0}{-1}
        \arrow{\xmax}{{\ymax/2}}{-1}{0}
        \arrow{\xmax}{{\ymax/2}}{-1}{1}
        \arrow{\xmax}{{\ymax/2}}{-1}{-1}
        % Top side
        \arrow{{\xmax/2}}{\ymax}{-1}{0}
        \arrow{{\xmax/2}}{\ymax}{1}{0}
        \arrow{{\xmax/2}}{\ymax}{0}{-1}
        \arrow{{\xmax/2}}{\ymax}{-1}{-1}
        \arrow{{\xmax/2}}{\ymax}{1}{-1}
        % Center point
        \arrow{{\xmax/2}}{{\ymax/2}}{-1}{-1}
        \arrow{{\xmax/2}}{{\ymax/2}}{-1}{0}
        \arrow{{\xmax/2}}{{\ymax/2}}{-1}{1}
        \arrow{{\xmax/2}}{{\ymax/2}}{0}{-1}
        \arrow{{\xmax/2}}{{\ymax/2}}{0}{1}
        \arrow{{\xmax/2}}{{\ymax/2}}{1}{-1}
        \arrow{{\xmax/2}}{{\ymax/2}}{1}{0}
        \arrow{{\xmax/2}}{{\ymax/2}}{1}{1}
        % Points
        \foreach \i in {0,...,2} {
            \foreach \j in {0,...,2} {
                \fill[orange] ({2*\i},{2*\j}) circle (4pt);
            }
        }
    \end{tikzpicture}}
    }
    \caption{Illustration of the refinement procedure. The orange points represent a subset of the initial grid. The blue points are those added during the refinement step. An arrow from an orange point to a blue point denotes that a chain is initialized at the blue point using the optimal mass function across all chains previously evaluated at the orange point.}
    \label{fig:refinement-schematic}
\end{figure}

Our goal is to determine the maximum and minimum merger rates as a function of the total abundance of PBH, $\fpbh$, and the fraction of those PBH that lie in the \DP~mass range, $\rdp$. Thus, in principle, we must perform the optimization described in the previous section at every point in this parameter space, independently. However, the optimization process is computationally expensive, so it cannot be applied directly to a fine grid in $(\rdp, \fpbh)$. Instead, we use the simulated annealing algorithm on a coarse grid, and then use an alternative technique to interpolate between the resulting optima.

First, we note that this interpolation process is not simply an aesthetic matter. In principle, a small displacement in the $(\rdp,\fpbh)$ plane can produce a sharp discontinuity in the shape of the optimal mass function, leading to discrete regions in which the optimal mass function evolves very differently with $\rdp$ and $\fpbh$. This is especially difficult to forecast when observational constraints are included. The situation is analogous to the behavior of the order parameter in a first-order phase transition: in this case, a small displacement in temperature discontinuously changes the location of the minimum of the free energy. In our case, rather than a sharp transition between two minima of the free energy, there could be a sharp transition between two shapes of the mass function. A naive interpolation of a coarse grid of points risks missing any such structure.

Therefore, we extend our coarse grid to a finer subgrid using the following procedure. First, each interval in the grid is halved to produce a refined grid. The initial mass functions for each new point on the grid are borrowed from its nearest neighbours: that is, for each neighbour, we run an independent chain at the new parameter point starting from the neighbour's optimal mass function across all of the neighbour's chains. One step of the process is illustrated in \cref{fig:refinement-schematic}. Even if there is a transition of the kind described above between the new point and some of its neighbours, it is still likely that the optimum at the new point is close in shape to that of at least one neighbour. Thus, one expects that only mild adjustment of these mass functions is needed to converge to the optimum at the new point.

Since we assume that at least one of the optimal mass functions drawn from the neighbouring points is close to the global optimum of the new point, there is no need for the variable temperature of simulated annealing: we need only locate the nearby optimum more precisely. We perform this adjustment by producing a chain of $10^5$ steps with the Metropolis--Hastings algorithm, which is structurally similar to simulated annealing, but with a fixed temperature. We perform the entire grid refinement procedure twice to obtain a sufficiently fine grid.

Finally, the optima from all points are `mixed' as follows. For each point on the refined grid, we generate another set of $10^5$-step Metropolis--Hastings chains, each with a different initial mass function. One chain begins with the optimal mass function from the point itself. Another chain is initialized from the optimal mass function of each nearest neighbour. The mixing process is performed four times, so an optimal mass function shape found at any point in a block between initial grid points can propagate to other points in the same block.

After the refinement and mixing processes are performed, the result is a non-trivial interpolation of the initial grid, which forms the basis for our results in the following sections.

\begin{figure*}\centering
    \includegraphics[width=0.9\textwidth]{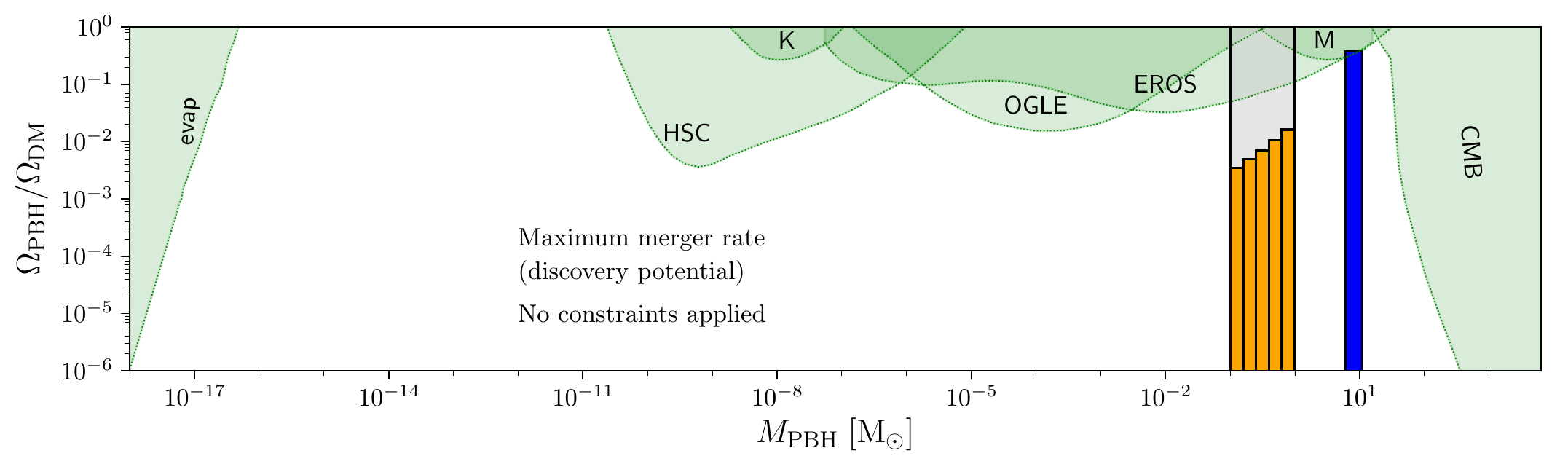}
    \\
    \includegraphics[width=0.9\textwidth]{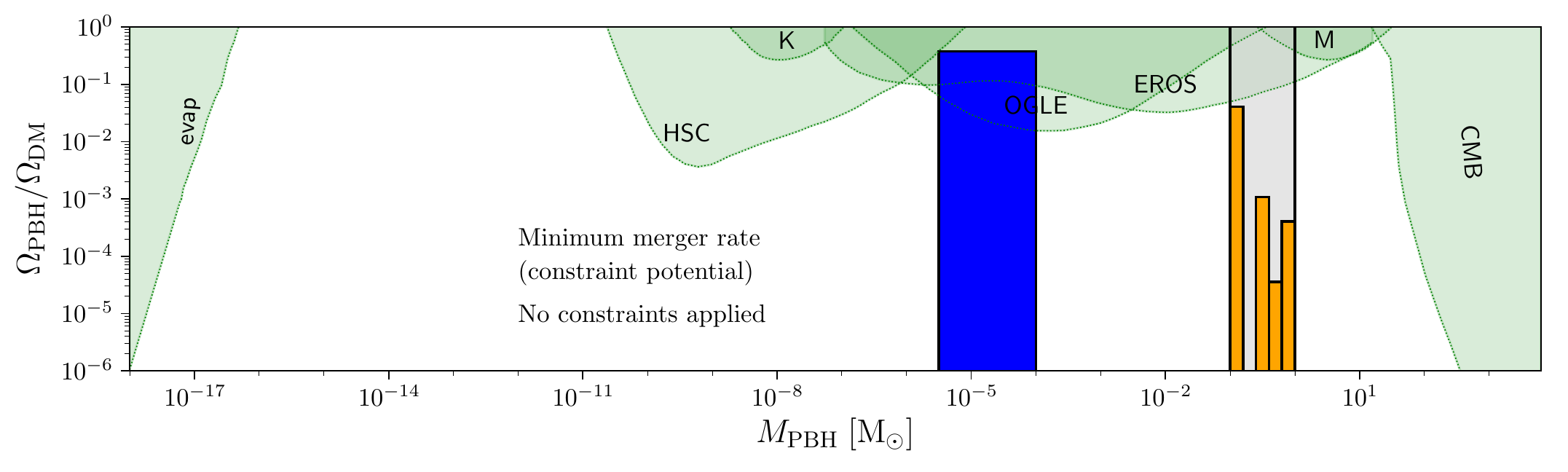}
    \caption{Optimal maximizing (top) and minimizing (bottom) mass functions with $\fpbh=0.5$ and $\rdp=0.1$ in the absence of observational constraints. Each mass function is shown as a set of discrete bars with height $\OmegaPBH(m_i)/\OmegaDM\equiv\psi(m_i)\Delta m_i$, i.e., the height of each bar indicates the total mass in the bin. The maximum merger rate corresponds to the most optimistic discovery potential, and the minimum merger rate to the most pessimistic constraint potential. The constraint curves are \emph{not} used to constrain the mass function, and are shown here only for reference and comparison with \cref{fig:constrained-optima}. The \DP~window is indicated by the shaded gray region, and the mass function is colored orange therein.
    \constraintlabels
    }
    \label{fig:free-optima}
\end{figure*}

\section{Results}
\label{sec:results}
We now examine the results of our numerical optimization. First we show results for individual parameter points, and compare the shapes of optimal mass functions to our analytical expectations. Then we show minimal and maximal merger rates with and without observational constraints.

\subsection{Shape of the mass function}
To understand the shapes of the mass functions that optimize the merger rate, we first neglect observational constraints to facilitate comparison with the analysis in \cref{sec:analytical}. \Cref{fig:free-optima} shows mass functions that minimize and maximize the merger rate without regard to observational constraints for the parameter point $(\rdp,\fpbh)=(0.1,0.5)$ (i.e. 50 per cent of DM is in PBH of any mass, and 10 per cent of the PBH density is accounted for by the \DP~window). The two mass functions are mostly distinguished by two features. First, they have clearly different behavior outside the \DP~window: the \DP~merger rate is enhanced when the remainder of the PBH are placed at a higher mass, above the top of the window, and it is reduced when they are placed at lower mass, below the bottom of the window. Secondly, as expected from our simplified analysis in \cref{sec:analytical}, the maximizer is broad within the \DP~window, while the minimizer is sharply peaked and multimodal.

Contrary to our naive expectation, the maximizer prefers higher masses within the \DP~window, while the minimizer prefers lower masses. This is because the full numerical calculation accounts for detectability, and the mergers of heavier black holes are detectable in a larger volume. This also accounts for the behavior of the mass function outside the \DP~window. Recall that the mergers of \DP~black holes with \emph{heavier} black holes are generally observable, and we assume that the lighter black hole is identifiably primordial in such a merger. However, the merger of a \DP~black hole with a \emph{lighter} black hole may not be observable, or may be observable only in such a small effective volume that our assumptions for calculating the merger rate are not valid. Thus, if the 90 per cent of PBH which lie outside the \DP~window are positioned at higher masses, the observable merger rate is enhanced.

\begin{figure*}\centering
    \includegraphics[width=0.9\textwidth]{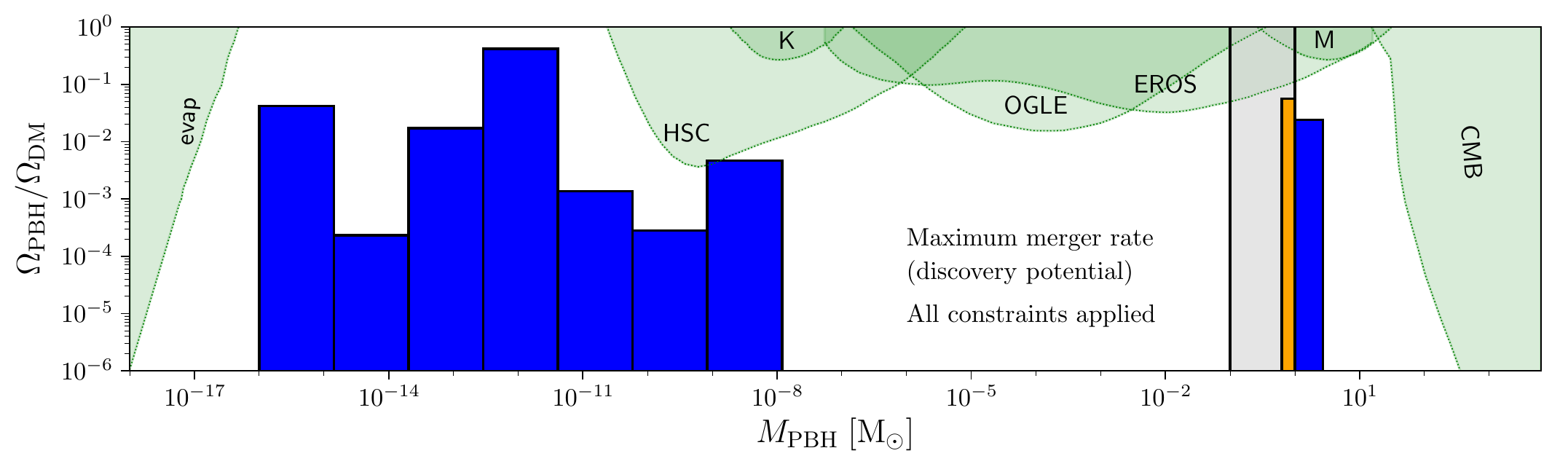}
    \\
    \includegraphics[width=0.9\textwidth]{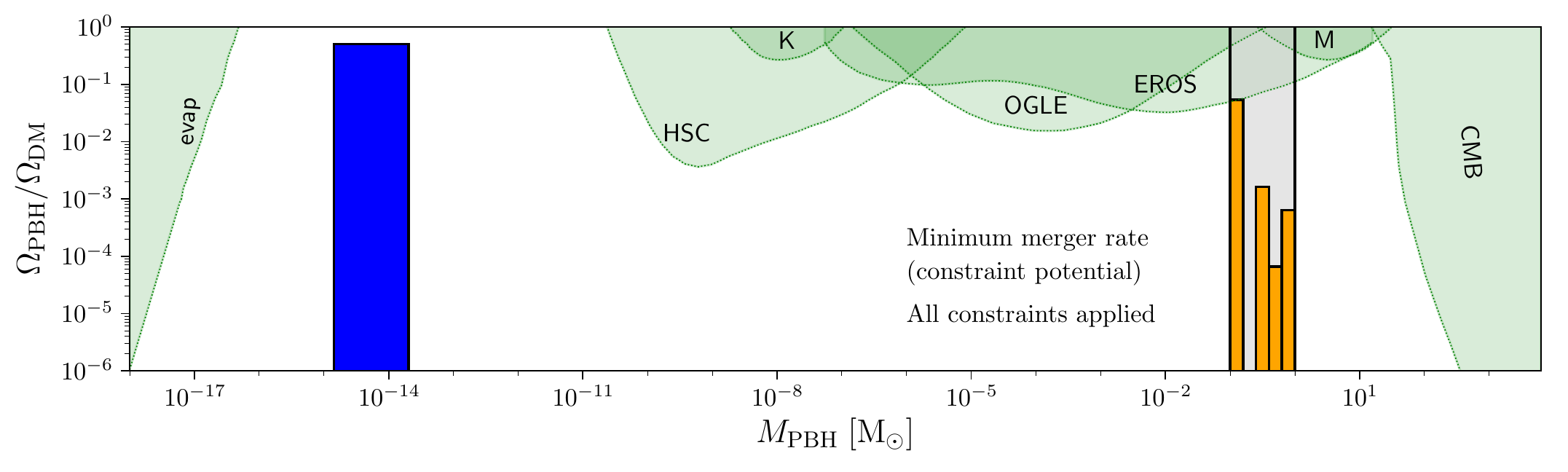}
    \caption{Optimal maximizing (top) and minimizing (bottom) mass functions with $\fpbh=0.5$ and $\rdp=0.1$. Each mass function is shown as a set of discrete bars with height $\OmegaPBH(m_i)/\OmegaDM\equiv\psi(m_i)\Delta m_i$, i.e., the height of each bar indicates the total mass in the bin. All observational constraints are applied. The maximum merger rate corresponds to the most optimistic discovery potential, and the minimum merger rate to the most pessimistic constraint potential. The \DP~window is indicated by the shaded gray region, and the mass function is coloured orange therein.
    \constraintlabels
    }
    \label{fig:constrained-optima}
\end{figure*}

Having noted the behavior of the optimal mass functions in the absence of observational constraints, we now turn to the results of constrained optimization in \cref{fig:constrained-optima}. The general features of these optima are similar to their unconstrained counterparts, and observational constraints modify the shapes of the optimal mass functions in a comprehensible manner. The maximal merger rate is still obtained with additional PBH positioned above the \DP~window, but observational bounds now strictly constrain the position of this peak. The mass function which minimizes the merger rate is not subject to strong constraints within the \DP~window, but the additional PBH at lower masses must now be repositioned to the mass range where constraints are weaker.

\subsection{Constraints and discovery prospects}
Minimum and maximum merger rates with all constraints applied are shown as a function of $\rdp$ and $\fpbh$ in \cref{fig:min-constrained,fig:max-constrained}. The minimum merger rate corresponds to LIGO's constraint potential: even given complete freedom in the mass function, there is no way to obtain a lower observable merger rate. The maximum merger rate corresponds to the discovery potential, i.e., the most optimistic scenario given any mass function.

The light gray region in the top right corner of each panel indicates parameter points where the numerical procedure was unable to locate any mass function consistent with constraints. This is the portion of parameter space that is already ruled out by other observables, including the SGWB. The extent of this region can be estimated using the semi-analytical procedure of \cite{Lehmann:2018ejc}, which can give the maximum allowed value of $\fpbh$ for fixed $\rdp$ if SGWB is neglected. This bound is the triangular dark gray region. Since the SGWB depends non-linearly on the mass function, it cannot be treated within the same semi-analytical framework. Thus, one expects the light gray region to extend slightly further than the dark gray region, which is exactly the behavior shown in \cref{fig:min-constrained,fig:max-constrained}.

Observe that there is a small gap between the minimum and maximum merger rates when $\rdp$ is near 1. This is simply because there is very limited freedom in the mass function under these conditions. On the other hand, when $\rdp\ll 1$, the minimum and maximum merger rates are radically different. In particular, while LIGO can only rule out mass functions with $\rdp\gtrsim0.1$, it can potentially discover PBH with only $\rdp\gtrsim10^{-4}$ with $\mathcal O(\SI{1}{\year})$ of data. The effect of observational constraints is evident from \cref{fig:constraint-impact}: in the absence of constraints, LIGO would potentially be sensitive to mergers of a subcomponent as small as $\rdp\sim10^{-6}$.

Finally, we note that the strength of the constraints is dependent on $\rdp$ and $\fpbh$ separately. One might expect the constraints to scale mainly with the product $\fdp=\rdp\fpbh$, i.e., the total abundance of black holes in the \DP~window. This is indeed the case for small values of $\fdp$. However, at larger values of $\fdp$, there are three effects that cause the constraints to depend on each of $\rdp$ and $\fpbh$ beyond their product. First, there is the uneven role of the observational constraints themselves. These have a complicated mass dependence, and thus introduce such dependence in the optimization results by limiting freedom in the mass function. This is mainly important at large $f_{\mathrm{\DP}}$. Secondly, there is the difference between the \DP~window for single PBHs and the $\mathrm\DP^2$ window for binaries: a PBH outside the $\mathrm{\DP}$ window can still contribute to the $\mathrm{\DP}^2$ merger rate by merging with a lighter PBH. Thirdly, even PBHs that do not participate in $\mathrm{\DP}^2$ mergers contribute to the formation and disruption of binaries in the $\mathrm{\DP}^2$ window. This holds true for both the merger rate of \citet{Chen:2018czv} and that of \citet{Raidal:2018bbj}.

\subsection{Convergence}
For an optimization problem of this kind, which is not generally convex, there is no reliable test of algorithmic convergence. In principle, it is always possible that the loss landscape has not been fully explored, and that in some corner, there is a point that outperforms the optima that we have discovered numerically. The best defense against this issue is to compare the numerical results against simplified analytical benchmarks, as we have carried out above.

However, we also perform two more direct tests of convergence. First, we have verified that we locate the global optimum in a low-dimensional example, where the features of the loss function can be analysed by inspection; and secondly, we perform a purely numerical test of convergence by comparing the results of many MCMC chains initialized in random configurations. We thus check directly that at benchmark points, all of our chains converge to the same merger rate within our fixed step count.

Numerical convergence is also supported qualitatively by comparison of nearby parameter points. Since we perform the optimization procedure on a grid of points in the $(\rdp,\fpbh)$ plane, nearest neighbours in this plane should converge to similar optima. Since the contours in \cref{fig:min-constrained,fig:max-constrained} are smooth, one might conclude that this constitutes evidence of convergence. However, note that in \cref{fig:min-constrained,fig:max-constrained}, optima from an initial run have been mixed between parameter points, as described in \cref{sec:numerics-refinement}. In particular, if a global optimum is discovered at only one point, it will subsequently propagate to the rest of the parameter space, even if chains originally produced elsewhere located very different optima. Thus, smoothness of the contours is only meaningful before mixing. Since the initial grid with random priors is relatively sparse, smoothness is difficult to assess quantitatively. However, we have verified that the qualitative features of the contours in \cref{fig:min-constrained,fig:max-constrained} are not affected by the mixing procedure, suggesting that each of the points in the initial grid is locating nearly the same optimum as that produced after mixing. Note that the sharp behavior at the top of \cref{fig:max-constrained} is entirely due to observational constraints, and disappears in their absence.

\section{Discussion and conclusions}
\label{sec:discussion}

The discovery of PBH would be a tremendous step forward in our understanding of cosmology. If PBH exist, they encode information about cosmic history in an epoch that we have yet to probe observationally. They also provide an empirical test of physics at extremely high scales and early times. Moreover, despite all observational constraints, PBH remain a viable and extremely simple candidate for cosmological dark matter.

Conveniently, any black hole with a mass below $\sim\SI{1}{M_\odot}$ cannot have an astrophysical origin. Gravitational wave observatories are well suited to identify black holes and to measure their masses precisely, so these instruments can detect a smoking-gun signature of the existence of PBH. Even one detection event involving a light black hole would provide unambiguous evidence for new physics. Subsequent exploration of the abundance and distribution of such black holes would test the possible formation scenarios, and potentially provide a direct handle on physics at very early times.

The problem lies in the interpretation of a null observational result. In principle, experimental results at LIGO constrain the population of light PBH, and in principle, again, LIGO may be sensitive to a very small abundance of such objects. However, both of these statements have a non-trivial dependence on the shape of the PBH mass function. Since the merger rate has a complicated non-linear dependence on the mass function, it is difficult to directly assess the significance of this uncertainty. In particular, the semi-analytical analysis of \cite{Lehmann:2018ejc} cannot accommodate the merger rate as a constraint on the PBH abundance.

\begin{figure}
    \includegraphics[width=\columnwidth]{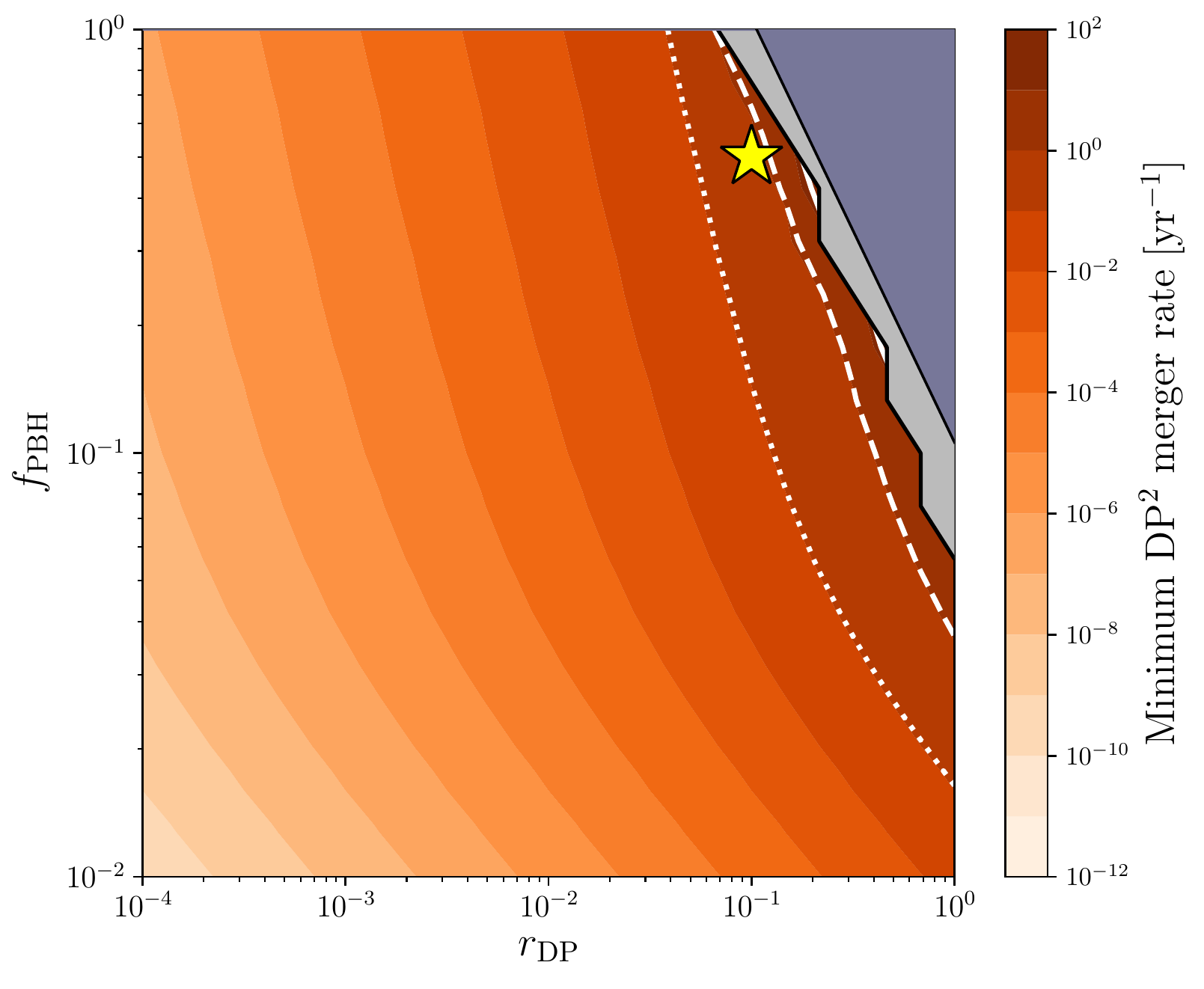}
    \caption{Minimum \DP~merger rate for mass functions constrained by all observables, including SGWB. The triangular region at the top right is ruled out by non-GW observables. The light region is ruled out by the combination of all observables. The solid, dashed, and dotted curves show contours with an \emph{observed} \DP~merger rate of 10, 1, and \SI{0.1}{\per\year}, respectively. The star ($\star$) indicates the point shown in the bottom panel of \cref{fig:constrained-optima}.}
    \label{fig:min-constrained}
\end{figure}
\begin{figure}
    \includegraphics[width=\columnwidth]{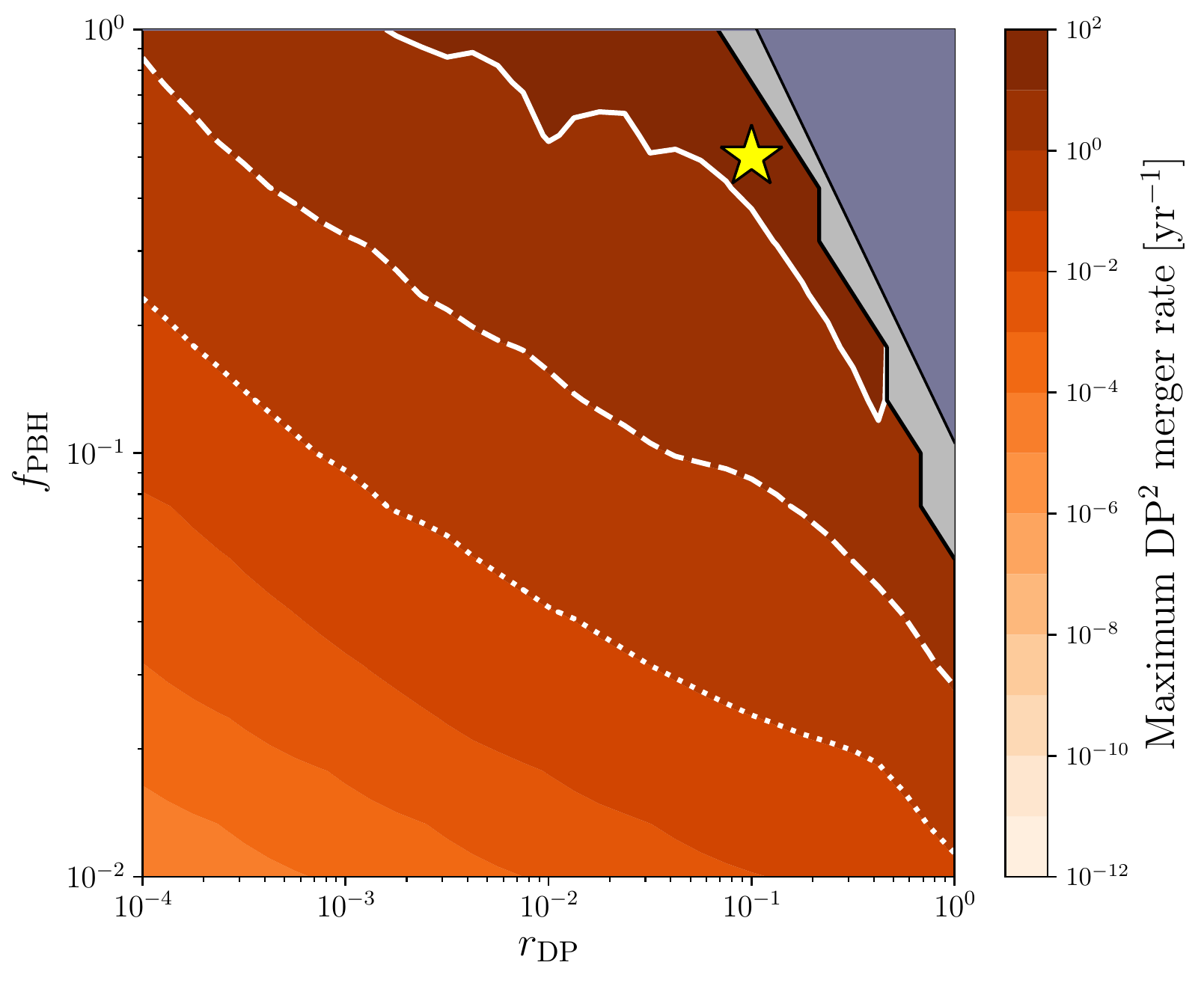}
    \caption{Maximum \DP~merger rate for mass functions constrained by all observables, including SGWB. The triangular region at the top right is ruled out by non-GW observables. The light region is ruled out by the combination of all observables. The solid, dashed, and dotted curves show contours with an \emph{observed} \DP~merger rate of 10, 1, and \SI{0.1}{\per\year}, respectively. The star ($\star$) indicates the point shown in the top panel of \cref{fig:constrained-optima}.}
    \label{fig:max-constrained}
\end{figure}

\begin{figure}
    \includegraphics[width=\columnwidth]{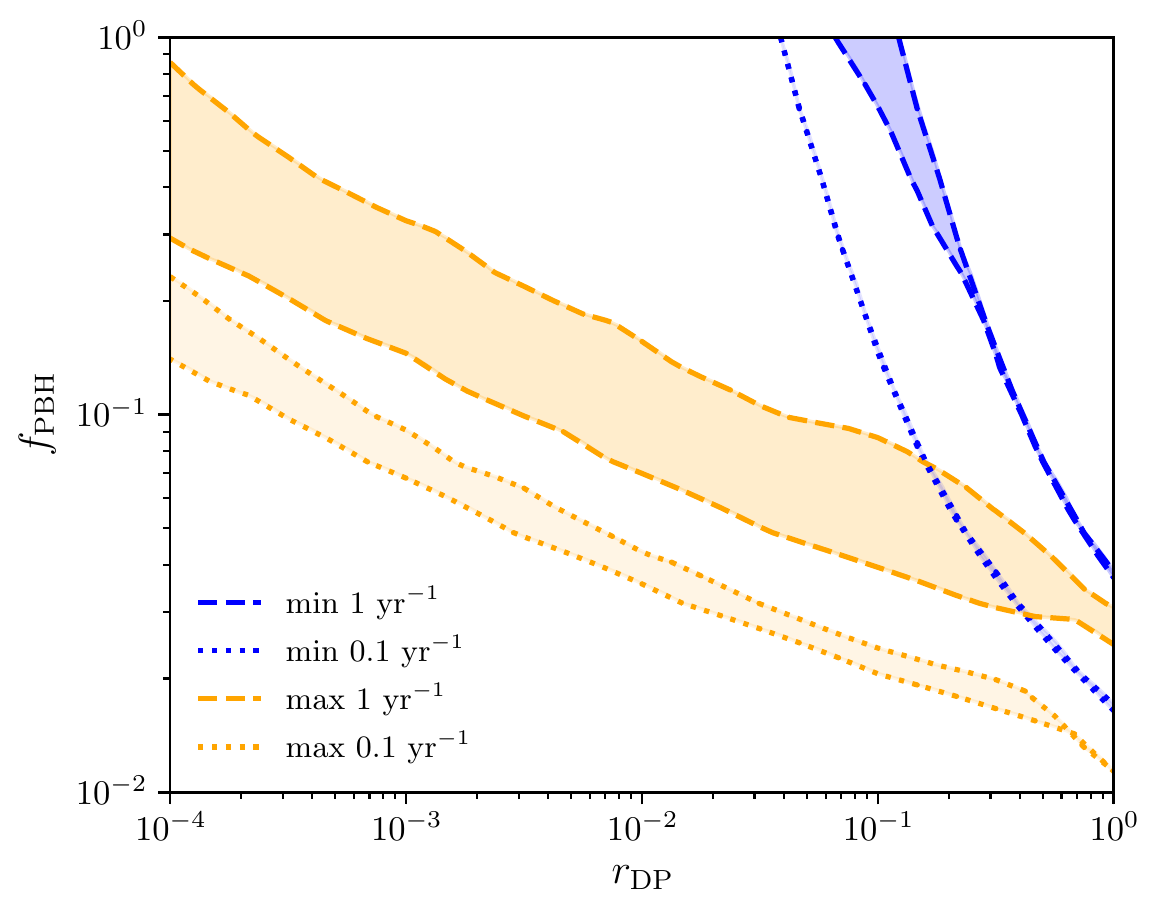}
    \caption{Contours with \DP~merger rate fixed to \SI{1}{\per\year} (dashed) and \SI{0.1}{\per\year} (dotted), with and without observational constraints. The bottom of each band shows the sensitivity with complete freedom in the mass function, and the top shows the sensitivity when observational constraints are included. The blue curves show the minimum merger rate, corresponding to the constraint potential. Note that the 10-yr minimum curves with and without constraints are essentially identical.}
    \label{fig:constraint-impact}
\end{figure}

In this work, we have quantified the uncertainty in the PBH merger rate that arises from freedom in the mass function, while accounting for observational constraints that restrict its shape. This uncertainty is reflected in the gap between the minimum-rate and maximum-rate contours in \cref{fig:constraint-impact}. While the two bands are not far apart for $\rdp\sim 1$, they are significantly different when $\rdp\ll 1$. Thus, it is necessary to consider the two contours as reflecting different notions of experimental sensitivity at LIGO. The minimal merger rate determines the extent of constraints that LIGO can set, if the mass function is allowed to vary freely. Conversely, the maximal merger rate determines the extent of the parameter space that can be probed by LIGO in the most optimistic scenario.

Our numerical results indicate that LIGO's constraint potential is limited to parameter space with $\rdp\gtrsim0.1$, and the prospects for improving this bound with binary black hole mergers are limited. On the other hand, LIGO's discovery potential extends as low as $\rdp\sim10^{-4}$, meaning that even a very small subcomponent of the PBH population that lies in our \DP~window can potentially yield a detection. This also establishes the relevance of constraints provided by other observables: in the absence of observational constraints, LIGO would be sensitive to $\rdp\sim10^{-6}$. Our results highlight the importance of evaluating detection prospects for specific PBH models using the full apparatus of the merger rate for extended mass functions---a small subcomponent of \DPBH~cannot be neglected.

One might wonder whether the optimal mass functions we consider in this work are realistic. Generally, there is good motivation to consider only specific forms of the mass function, particularly monochromatic, lognormal, or power-law shapes. However, most of the behavior that characterizes our optimal mass functions is captured by doubly or triply peaked mass functions, and note that a population of PBH with a multimodal mass function can easily be generated after inflation \citep{Carr:2018poi}. Thus, while the exact form of our optimal mass functions might require fine-tuning of initial conditions, approximate forms that retain a high or low merger rate are much more generic. The non-trivial requirement is that a peak should fall near the \DP~window to maximize discovery prospects. As yet, there is no direct evidence for such placement, but only circumstantial evidence from the distribution of mergers observed thus far.

Our results are inherently subject to theoretical uncertainties in the computation of the merger rate. While the form of the merger rate employed here reflects one of the most comprehensive estimates currently available, such formulae are best suited only to computations at the order of magnitude level. For instance, one potential issue in the rate calculation is the effect of other black holes in disrupting the formation of a binary. In our calculation, as discussed by \cite{Chen:2018czv}, we assume that two black holes of mass $m_i$ and $m_j$ do not form a binary if another black hole of mass $m_k\geq\min\{m_i,m_j\}$ is present in the volume between them. However, even if this were always the case, it is also possible that somewhat lighter black holes would have a similar effect. This would provide a mechanism for suppression of the merger rate, reducing the discovery potential and weakening the constraint we draw in this work.

As a cross-check, we have also computed the merger rate for each of our optimal mass functions using the formalism of \citet{Raidal:2018bbj}. Here, the influence of perturbing black holes on the binary is calculated by an entirely different method, as discussed in \cref{sec:merger-rate}. For our optimal mass functions, the merger rates obtained in each of the two formalisms are comparable, generally differing by an $\mathcal O(1)$ factor, but the difference can be as large as $\mathcal O(10)$ for some points. On the one hand, this is quite good agreement, given that these are two structurally different calculations with many inherent uncertainties, applied to complicated mass functions which differ substantially from standard benchmarks. On the other hand, the disagreement requires that we limit the interpretation of our results to the order-of-magnitude level.

Along similar lines, \citet{Jedamzik:2020ypm} recently showed numerically that including all subsequent three-body encounters after binary formation can significantly reduce the merger rate. The suppression described in that work can be as small as a $\mathcal O(2\text{--}20)$ factor, or as large as a $\mathcal O(10^3)$ factor, depending on the clustering properties of PBH. We thus consider a reduction of our calculated merger rate by at least a factor of $\mathcal O(10)$ to be physically well motivated. Thus, even at the order-of-magnitude level, it is possible that we overestimate the merger rate somewhat.

In light of the differences between \citet{Chen:2018czv} and \citet{Raidal:2018bbj}, and the uncertainties suggested by \citet{Jedamzik:2020ypm}, it is clear that any qualitative interpretation must include these substantial systematics. We therefore include contours with a merger rate of \SI{10}{\per\year} in \cref{fig:min-constrained,fig:max-constrained}. In this case, LIGO's discovery potential is reduced to $\rdp\gtrsim10^{-3}$, and constraint potential is lost completely: the \SI{10}{\per\year} contour in \cref{fig:min-constrained} is covered almost entirely by the existing non-merger constraints. Note, however, that if mergers of binaries formed in the early Universe are suppressed, binaries formed in the late Universe may make an important contribution to the rate, particularly if the density contrast in the late Universe is larger than expected. Ultimately, barring extreme modifications to the merger rate, our qualitative results stand. In particular, the gap between the maximal and minimal merger rates is very large at small $\rdp$, and is robust to adjustments in the calculation of the merger rate. However, further refinement in the prediction of the merger rate is certainly motivated.

In this work, we have focused on the direct observation of \DP~black holes as a smoking gun of the primordial-origin scenario. In the absence of such a direct signature, the SGWB associated with mergers over cosmic time may provide an additional probe of the PBH abundance. We do not evaluate SGWB as a discovery mechanism simply because such a detection would not constitute unambiguous evidence of new physics. It is possible that features of the SGWB may be connected to the features of the PBH population with enough precision to empirically test specific models, but since other physical mechanisms might also contribute to the SGWB, significant additional work would be required to confirm the existence of a population of PBH. However, we emphasize that the SGWB is still a sensitive probe of black holes in the \DP~window. In particular, mass function shapes that greatly enhance the merger rate can be ruled out by SGWB limits. In our framework, while we do not examine the SGWB as a tool for discovery, we do consistently include this observable as one of our constraints on the mass function: all of the constrained optima we consider, including those with maximal merger rates, are compatible with existing SGWB constraints. Significant improvement of observational bounds on the SGWB might limit freedom in the mass function, and might thus limit the discovery potential we infer in this work.

Our results show that while LIGO has limited power to constrain the abundance of light PBH, it nonetheless has significant discovery potential. The major obstruction to such sensitivity is not the sensitivity of the LIGO instrument, but the analysis pipeline. There are significant computational costs to conducting searches for mergers of light black holes, as discussed extensively by \cite{Magee:2018opb}, and these costs increase further if one searches for mergers of light black holes with heavier black holes. However, the freedom in the mass function and the associated uncertainty in the merger rate provides ample motivation for the refinement of methods for such searches, and even for the dedication of additional computational resources. A single observation of this type would have immense value, and gravitational wave observatories are in a unique position to make such a discovery.

\section*{Acknowledgements}
BVL and SP are partially supported  by the U.S. Department of Energy grant number DE-SC0010107. We acknowledge valuable conversations with Nergis Mavalvala and Sarah Shandera, and we thank Krzysztof Belczynski, Karsten Jedamzik, and Antonio Riotto for helpful comments on an earlier version of this work. This work has used the Hummingbird Computational Cluster operated by UC Santa Cruz Research Computing, and we thank the Research Computing staff for valuable support.

\section*{Data Availability}
The data underlying this article will be shared on reasonable request to the corresponding author.

\bibliographystyle{mnras}
\bibliography{main}

\begin{thebibliography}{}
\makeatletter
\relax
\def\mn@urlcharsother{\let\do\@makeother \do\$\do\&\do\#\do\^\do\_\do\%\do\~}
\def\mn@doi{\begingroup\mn@urlcharsother \@ifnextchar [ {\mn@doi@}
  {\mn@doi@[]}}
\def\mn@doi@[#1]#2{\def\@tempa{#1}\ifx\@tempa\@empty \href
  {http://dx.doi.org/#2} {doi:#2}\else \href {http://dx.doi.org/#2} {#1}\fi
  \endgroup}
\def\mn@eprint#1#2{\mn@eprint@#1:#2::\@nil}
\def\mn@eprint@arXiv#1{\href {http://arxiv.org/abs/#1} {{\tt arXiv:#1}}}
\def\mn@eprint@dblp#1{\href {http://dblp.uni-trier.de/rec/bibtex/#1.xml}
  {dblp:#1}}
\def\mn@eprint@#1:#2:#3:#4\@nil{\def\@tempa {#1}\def\@tempb {#2}\def\@tempc
  {#3}\ifx \@tempc \@empty \let \@tempc \@tempb \let \@tempb \@tempa \fi \ifx
  \@tempb \@empty \def\@tempb {arXiv}\fi \@ifundefined
  {mn@eprint@\@tempb}{\@tempb:\@tempc}{\expandafter \expandafter \csname
  mn@eprint@\@tempb\endcsname \expandafter{\@tempc}}}

\bibitem[\protect\citeauthoryear{Abbott et~al.}{Abbott
  et~al.}{2005}]{Abbott:2005pf}
Abbott B.,  et~al., 2005, \mn@doi [Phys. Rev.] {10.1103/PhysRevD.72.082002},
  D72, 082002

\bibitem[\protect\citeauthoryear{Abbott et~al.}{Abbott
  et~al.}{2009}]{Abbott:2009ws}
Abbott B.,  et~al., 2009, \mn@doi [Nature] {10.1038/nature08278}, 460, 990

\bibitem[\protect\citeauthoryear{Abbott et~al.}{Abbott
  et~al.}{2016a}]{TheLIGOScientific:2016pea}
Abbott B.,  et~al., 2016a, \mn@doi [Phys. Rev. X] {10.1103/PhysRevX.6.041015},
  6, 041015

\bibitem[\protect\citeauthoryear{Abbott et~al.}{Abbott
  et~al.}{2016b}]{Abbott:2016blz}
Abbott B.,  et~al., 2016b, \mn@doi [Phys. Rev. Lett.]
  {10.1103/PhysRevLett.116.061102}, 116, 061102

\bibitem[\protect\citeauthoryear{Abbott et~al.}{Abbott
  et~al.}{2016c}]{TheLIGOScientific:2016src}
Abbott B.,  et~al., 2016c, \mn@doi [Phys. Rev. Lett.]
  {10.1103/PhysRevLett.116.221101}, 116, 221101

\bibitem[\protect\citeauthoryear{Abbott et~al.}{Abbott
  et~al.}{2016d}]{Abbott:2016nmj}
Abbott B.~P.,  et~al., 2016d, \mn@doi [Phys. Rev. Lett.]
  {10.1103/PhysRevLett.116.241103}, 116, 241103

\bibitem[\protect\citeauthoryear{Abbott et~al.}{Abbott
  et~al.}{2017a}]{TheLIGOScientific:2016dpb}
Abbott B.~P.,  et~al., 2017a, \mn@doi [Phys. Rev. Lett.]
  {10.1103/PhysRevLett.118.121101}, 118, 121101

\bibitem[\protect\citeauthoryear{Abbott et~al.}{Abbott
  et~al.}{2017b}]{Abbott:2017vtc}
Abbott B.~P.,  et~al., 2017b, \mn@doi [Phys. Rev. Lett.]
  {10.1103/PhysRevLett.118.221101}, 118, 221101

\bibitem[\protect\citeauthoryear{Abbott et~al.}{Abbott
  et~al.}{2017c}]{Abbott:2017oio}
Abbott B.,  et~al., 2017c, \mn@doi [Phys. Rev. Lett.]
  {10.1103/PhysRevLett.119.141101}, 119, 141101

\bibitem[\protect\citeauthoryear{Abbott et~al.}{Abbott
  et~al.}{2017d}]{TheLIGOScientific:2017qsa}
Abbott B.,  et~al., 2017d, \mn@doi [Phys. Rev. Lett.]
  {10.1103/PhysRevLett.119.161101}, 119, 161101

\bibitem[\protect\citeauthoryear{Abbott et~al.}{Abbott
  et~al.}{2017e}]{GBM:2017lvd}
Abbott B.,  et~al., 2017e, \mn@doi [Astrophys. J. Lett.]
  {10.3847/2041-8213/aa91c9}, 848, L12

\bibitem[\protect\citeauthoryear{Abbott et~al.}{Abbott
  et~al.}{2017f}]{Monitor:2017mdv}
Abbott B.,  et~al., 2017f, \mn@doi [Astrophys. J. Lett.]
  {10.3847/2041-8213/aa920c}, 848, L13

\bibitem[\protect\citeauthoryear{Abbott et~al.}{Abbott
  et~al.}{2017g}]{Abbott:2017gyy}
Abbott B.~P.,  et~al., 2017g, \mn@doi [Astrophys. J.]
  {10.3847/2041-8213/aa9f0c}, 851, L35

\bibitem[\protect\citeauthoryear{Abbott et~al.}{Abbott
  et~al.}{2019a}]{LIGOScientific:2018mvr}
Abbott B.,  et~al., 2019a, \mn@doi [Phys. Rev. X] {10.1103/PhysRevX.9.031040},
  9, 031040

\bibitem[\protect\citeauthoryear{Abbott et~al.}{Abbott
  et~al.}{2019b}]{Authors:2019qbw}
Abbott B.,  et~al., 2019b, \mn@doi [Phys. Rev. Lett.]
  {10.1103/PhysRevLett.123.161102}, 123, 161102

\bibitem[\protect\citeauthoryear{Abbott et~al.}{Abbott
  et~al.}{2020a}]{LIGOScientific:2020stg}
Abbott R.,  et~al., 2020a, \mn@doi [Phys. Rev. D]
  {10.1103/PhysRevD.102.043015}, 102, 043015

\bibitem[\protect\citeauthoryear{Abbott et~al.}{Abbott
  et~al.}{2020b}]{Abbott:2020uma}
Abbott B.,  et~al., 2020b, \mn@doi [Astrophys. J. Lett.]
  {10.3847/2041-8213/ab75f5}, 892, L3

\bibitem[\protect\citeauthoryear{Abbott et~al.}{Abbott
  et~al.}{2020c}]{Abbott:2020khf}
Abbott R.,  et~al., 2020c, \mn@doi [Astrophys. J.] {10.3847/2041-8213/ab960f},
  896, L44

\bibitem[\protect\citeauthoryear{Ali-Haïmoud \& Kamionkowski}{Ali-Haïmoud \&
  Kamionkowski}{2017}]{Ali-Haimoud:2016mbv}
Ali-Haïmoud Y.,  Kamionkowski M.,  2017, \mn@doi [Phys. Rev. D]
  {10.1103/PhysRevD.95.043534}, 95, 043534

\bibitem[\protect\citeauthoryear{Allsman et~al.}{Allsman
  et~al.}{2001}]{Allsman:2000kg}
Allsman R.,  et~al., 2001, \mn@doi [Astrophys. J. Lett.] {10.1086/319636}, 550,
  L169

\bibitem[\protect\citeauthoryear{Bailyn, Jain, Coppi  \& Orosz}{Bailyn
  et~al.}{1998}]{Bailyn:1997xt}
Bailyn C.~D.,  Jain R.~K.,  Coppi P.,   Orosz J.~A.,  1998, \mn@doi [Astrophys.
  J.] {10.1086/305614}, 499, 367

\bibitem[\protect\citeauthoryear{Belczynski, Bulik, Fryer, Ruiter, Vink  \&
  Hurley}{Belczynski et~al.}{2010}]{Belczynski:2009xy}
Belczynski K.,  Bulik T.,  Fryer C.~L.,  Ruiter A.,  Vink J.~S.,   Hurley
  J.~R.,  2010, \mn@doi [Astrophys. J.] {10.1088/0004-637X/714/2/1217}, 714,
  1217

\bibitem[\protect\citeauthoryear{Belczynski et~al.}{Belczynski
  et~al.}{2020}]{Belczynski:2017gds}
Belczynski K.,  et~al., 2020, \mn@doi [Astron. Astrophys.]
  {10.1051/0004-6361/201936528}, 636, A104

\bibitem[\protect\citeauthoryear{Bird, Cholis, Muñoz, Ali-Haïmoud,
  Kamionkowski, Kovetz, Raccanelli  \& Riess}{Bird et~al.}{2016}]{Bird:2016dcv}
Bird S.,  Cholis I.,  Muñoz J.~B.,  Ali-Haïmoud Y.,  Kamionkowski M.,  Kovetz
  E.~D.,  Raccanelli A.,   Riess A.~G.,  2016, \mn@doi [Phys. Rev. Lett.]
  {10.1103/PhysRevLett.116.201301}, 116, 201301

\bibitem[\protect\citeauthoryear{Brandt}{Brandt}{2016}]{Brandt:2016aco}
Brandt T.~D.,  2016, \mn@doi [Astrophys. J. Lett.]
  {10.3847/2041-8205/824/2/L31}, 824, L31

\bibitem[\protect\citeauthoryear{Calmet, Carr  \& Winstanley}{Calmet
  et~al.}{2014}]{Calmet:2014dea}
Calmet X.,  Carr B.,   Winstanley E.,  2014, {Quantum Black Holes}.
SpringerBriefs in Physics, Springer, Berlin, \mn@doi{10.1007/978-3-642-38939-9}

\bibitem[\protect\citeauthoryear{Capela, Pshirkov  \& Tinyakov}{Capela
  et~al.}{2013}]{Capela:2013yf}
Capela F.,  Pshirkov M.,   Tinyakov P.,  2013, \mn@doi [Phys. Rev. D]
  {10.1103/PhysRevD.87.123524}, 87, 123524

\bibitem[\protect\citeauthoryear{Carr}{Carr}{2003}]{Carr:2003bj}
Carr B.~J.,  2003, \mn@doi [Lect. Notes Phys.] {10.1007/978-3-540-45230-0\_7},
  631, 301

\bibitem[\protect\citeauthoryear{Carr \& Kuhnel}{Carr \&
  Kuhnel}{2019}]{Carr:2018poi}
Carr B.,  Kuhnel F.,  2019, \mn@doi [Phys. Rev. D]
  {10.1103/PhysRevD.99.103535}, 99, 103535

\bibitem[\protect\citeauthoryear{Carr \& Kuhnel}{Carr \&
  Kuhnel}{2020}]{Carr:2020xqk}
Carr B.,  Kuhnel F.,  2020, \mn@doi [Ann. Rev. Nucl. Part. Sci.]
  {10.1146/annurev-nucl-050520-125911}, 70, 355

\bibitem[\protect\citeauthoryear{Carr, Kohri, Sendouda  \& Yokoyama}{Carr
  et~al.}{2010}]{Carr:2009jm}
Carr B.,  Kohri K.,  Sendouda Y.,   Yokoyama J.,  2010, \mn@doi [Phys. Rev. D]
  {10.1103/PhysRevD.81.104019}, 81, 104019

\bibitem[\protect\citeauthoryear{Carr, Kuhnel  \& Sandstad}{Carr
  et~al.}{2016}]{Carr:2016drx}
Carr B.,  Kuhnel F.,   Sandstad M.,  2016, \mn@doi [Phys. Rev. D]
  {10.1103/PhysRevD.94.083504}, 94, 083504

\bibitem[\protect\citeauthoryear{Carr, Raidal, Tenkanen, Vaskonen  \&
  Veermäe}{Carr et~al.}{2017}]{Carr:2017jsz}
Carr B.,  Raidal M.,  Tenkanen T.,  Vaskonen V.,   Veermäe H.,  2017, \mn@doi
  [Phys. Rev. D] {10.1103/PhysRevD.96.023514}, 96, 023514

\bibitem[\protect\citeauthoryear{Carr, Kohri, Sendouda  \& Yokoyama}{Carr
  et~al.}{2020}]{Carr:2020gox}
Carr B.,  Kohri K.,  Sendouda Y.,   Yokoyama J.,  2020, preprint
  (\href{https://arxiv.org/abs/2002.12778}{arXiv:2002.12778})

\bibitem[\protect\citeauthoryear{Chandrasekhar}{Chandrasekhar}{1931}]{Chandrasekhar:1931ih}
Chandrasekhar S.,  1931, \mn@doi [Astrophys. J.] {10.1086/143324}, 74, 81

\bibitem[\protect\citeauthoryear{Chapline}{Chapline}{1975}]{Chapline:1975ojl}
Chapline G.~F.,  1975, \mn@doi [Nature] {10.1038/253251a0}, 253, 251

\bibitem[\protect\citeauthoryear{Chen \& Huang}{Chen \&
  Huang}{2018}]{Chen:2018czv}
Chen Z.-C.,  Huang Q.-G.,  2018, \mn@doi [Astrophys. J.]
  {10.3847/1538-4357/aad6e2}, 864, 61

\bibitem[\protect\citeauthoryear{Chen \& Huang}{Chen \&
  Huang}{2020}]{Chen:2019irf}
Chen Z.-C.,  Huang Q.-G.,  2020, \mn@doi [JCAP]
  {10.1088/1475-7516/2020/08/039}, 08, 039

\bibitem[\protect\citeauthoryear{Cholis, Kovetz, Ali-Haïmoud, Bird,
  Kamionkowski, Muñoz  \& Raccanelli}{Cholis et~al.}{2016}]{Cholis:2016kqi}
Cholis I.,  Kovetz E.~D.,  Ali-Haïmoud Y.,  Bird S.,  Kamionkowski M.,  Muñoz
  J.~B.,   Raccanelli A.,  2016, \mn@doi [Phys. Rev. D]
  {10.1103/PhysRevD.94.084013}, 94, 084013

\bibitem[\protect\citeauthoryear{Clesse \& García-Bellido}{Clesse \&
  García-Bellido}{2017}]{Clesse:2016ajp}
Clesse S.,  García-Bellido J.,  2017, \mn@doi [Phys. Dark Univ.]
  {10.1016/j.dark.2017.10.001}, 18, 105

\bibitem[\protect\citeauthoryear{Coughlin \& Dietrich}{Coughlin \&
  Dietrich}{2019}]{Coughlin:2019kqf}
Coughlin M.~W.,  Dietrich T.,  2019, \mn@doi [Phys. Rev. D]
  {10.1103/PhysRevD.100.043011}, 100, 043011

\bibitem[\protect\citeauthoryear{De~Luca, Franciolini, Pani  \& Riotto}{De~Luca
  et~al.}{2020a}]{DeLuca:2020bjf}
De~Luca V.,  Franciolini G.,  Pani P.,   Riotto A.,  2020a, \mn@doi [JCAP]
  {10.1088/1475-7516/2020/04/052}, 04, 052

\bibitem[\protect\citeauthoryear{De~Luca, Franciolini, Pani  \& Riotto}{De~Luca
  et~al.}{2020b}]{DeLuca:2020qqa}
De~Luca V.,  Franciolini G.,  Pani P.,   Riotto A.,  2020b, \mn@doi [JCAP]
  {10.1088/1475-7516/2020/06/044}, 06, 044

\bibitem[\protect\citeauthoryear{Dolgov, Kuranov, Mitichkin, Porey, Postnov,
  Sazhina  \& Simkin}{Dolgov et~al.}{2020}]{Dolgov:2020xzo}
Dolgov A.~D.,  Kuranov A.~G.,  Mitichkin N.~A.,  Porey S.,  Postnov K.~A.,
  Sazhina O.~S.,   Simkin I.~V.,  2020, \mn@doi [JCAP]
  {10.1088/1475-7516/2020/12/017}, 12, 017

\bibitem[\protect\citeauthoryear{Farr, Sravan, Cantrell, Kreidberg, Bailyn,
  Mandel  \& Kalogera}{Farr et~al.}{2011}]{Farr:2010tu}
Farr W.~M.,  Sravan N.,  Cantrell A.,  Kreidberg L.,  Bailyn C.~D.,  Mandel I.,
    Kalogera V.,  2011, \mn@doi [Astrophys. J.] {10.1088/0004-637X/741/2/103},
  741, 103

\bibitem[\protect\citeauthoryear{Farr, Holz  \& Farr}{Farr
  et~al.}{2018}]{Farr:2017gtv}
Farr B.,  Holz D.~E.,   Farr W.~M.,  2018, \mn@doi [Astrophys. J. Lett.]
  {10.3847/2041-8213/aaaa64}, 854, L9

\bibitem[\protect\citeauthoryear{Fernandez \& Profumo}{Fernandez \&
  Profumo}{2019}]{Fernandez:2019kyb}
Fernandez N.,  Profumo S.,  2019, \mn@doi [JCAP]
  {10.1088/1475-7516/2019/08/022}, 08, 022

\bibitem[\protect\citeauthoryear{Gerosa \& Berti}{Gerosa \&
  Berti}{2017}]{Gerosa:2017kvu}
Gerosa D.,  Berti E.,  2017, \mn@doi [Phys. Rev. D]
  {10.1103/PhysRevD.95.124046}, 95, 124046

\bibitem[\protect\citeauthoryear{Graham, Rajendran  \& Varela}{Graham
  et~al.}{2015}]{Graham:2015apa}
Graham P.~W.,  Rajendran S.,   Varela J.,  2015, \mn@doi [Phys. Rev. D]
  {10.1103/PhysRevD.92.063007}, 92, 063007

\bibitem[\protect\citeauthoryear{Griest, Cieplak  \& Lehner}{Griest
  et~al.}{2014}]{Griest:2013aaa}
Griest K.,  Cieplak A.~M.,   Lehner M.~J.,  2014, \mn@doi [Astrophys. J.]
  {10.1088/0004-637X/786/2/158}, 786, 158

\bibitem[\protect\citeauthoryear{Hassall}{Hassall}{1904}]{Hassall:1904}
Hassall J.,  1904, The Old Nursery Stories and Rhymes.
Blackie \& Son

\bibitem[\protect\citeauthoryear{Hastings}{Hastings}{1970}]{Hastings:1970aa}
Hastings W.,  1970, \mn@doi [Biometrika] {10.1093/biomet/57.1.97}, 57, 97

\bibitem[\protect\citeauthoryear{Hawking}{Hawking}{1971}]{10.1093/mnras/152.1.75}
Hawking S.,  1971, \mn@doi [Monthly Notices of the Royal Astronomical Society]
  {10.1093/mnras/152.1.75}, 152, 75

\bibitem[\protect\citeauthoryear{Inomata, Kawasaki, Mukaida  \&
  Yanagida}{Inomata et~al.}{2018}]{Inomata:2017vxo}
Inomata K.,  Kawasaki M.,  Mukaida K.,   Yanagida T.~T.,  2018, \mn@doi [Phys.
  Rev. D] {10.1103/PhysRevD.97.043514}, 97, 043514

\bibitem[\protect\citeauthoryear{Jedamzik}{Jedamzik}{2020}]{Jedamzik:2020ypm}
Jedamzik K.,  2020, \mn@doi [JCAP] {10.1088/1475-7516/2020/09/022}, 09, 022

\bibitem[\protect\citeauthoryear{Katz, Kopp, Sibiryakov  \& Xue}{Katz
  et~al.}{2018}]{Katz:2018zrn}
Katz A.,  Kopp J.,  Sibiryakov S.,   Xue W.,  2018, \mn@doi [JCAP]
  {10.1088/1475-7516/2018/12/005}, 12, 005

\bibitem[\protect\citeauthoryear{Khlopov}{Khlopov}{2010}]{Khlopov:2008qy}
Khlopov M.~Y.,  2010, \mn@doi [Res. Astron. Astrophys.]
  {10.1088/1674-4527/10/6/001}, 10, 495

\bibitem[\protect\citeauthoryear{Kirkpatrick, Gelatt  \& Vecchi}{Kirkpatrick
  et~al.}{1983}]{Kirkpatrick:1983zz}
Kirkpatrick S.,  Gelatt C.,   Vecchi M.,  1983, \mn@doi [Science]
  {10.1126/science.220.4598.671}, 220, 671

\bibitem[\protect\citeauthoryear{Koushiappas \& Loeb}{Koushiappas \&
  Loeb}{2017}]{Koushiappas:2017chw}
Koushiappas S.~M.,  Loeb A.,  2017, \mn@doi [Phys. Rev. Lett.]
  {10.1103/PhysRevLett.119.041102}, 119, 041102

\bibitem[\protect\citeauthoryear{Lehmann, Profumo  \& Yant}{Lehmann
  et~al.}{2018}]{Lehmann:2018ejc}
Lehmann B.~V.,  Profumo S.,   Yant J.,  2018, \mn@doi [JCAP]
  {10.1088/1475-7516/2018/04/007}, 1804, 007

\bibitem[\protect\citeauthoryear{Magee et~al.,}{Magee
  et~al.}{2018}]{Magee:2018opb}
Magee R.,  et~al., 2018, \mn@doi [Phys. Rev.] {10.1103/PhysRevD.98.103024},
  D98, 103024

\bibitem[\protect\citeauthoryear{Mandic, Bird  \& Cholis}{Mandic
  et~al.}{2016}]{Mandic:2016lcn}
Mandic V.,  Bird S.,   Cholis I.,  2016, \mn@doi [Phys. Rev. Lett.]
  {10.1103/PhysRevLett.117.201102}, 117, 201102

\bibitem[\protect\citeauthoryear{Metropolis, Rosenbluth, Rosenbluth, Teller  \&
  Teller}{Metropolis et~al.}{1953}]{Metropolis:1953am}
Metropolis N.,  Rosenbluth A.,  Rosenbluth M.,  Teller A.,   Teller E.,  1953,
  \mn@doi [J. Chem. Phys.] {10.1063/1.1699114}, 21, 1087

\bibitem[\protect\citeauthoryear{Monroy-Rodríguez \& Allen}{Monroy-Rodríguez
  \& Allen}{2014}]{Monroy-Rodriguez:2014ula}
Monroy-Rodríguez M.~A.,  Allen C.,  2014, \mn@doi [Astrophys. J.]
  {10.1088/0004-637X/790/2/159}, 790, 159

\bibitem[\protect\citeauthoryear{Montero-Camacho, Fang, Vasquez, Silva  \&
  Hirata}{Montero-Camacho et~al.}{2019}]{Montero-Camacho:2019jte}
Montero-Camacho P.,  Fang X.,  Vasquez G.,  Silva M.,   Hirata C.~M.,  2019,
  \mn@doi [JCAP] {10.1088/1475-7516/2019/08/031}, 08, 031

\bibitem[\protect\citeauthoryear{Niikura et~al.}{Niikura
  et~al.}{2019}]{Niikura:2017zjd}
Niikura H.,  et~al., 2019, \mn@doi [Nature Astron.]
  {10.1038/s41550-019-0723-1}, 3, 524

\bibitem[\protect\citeauthoryear{Ozel, Psaltis, Narayan  \& McClintock}{Ozel
  et~al.}{2010}]{Ozel:2010su}
Ozel F.,  Psaltis D.,  Narayan R.,   McClintock J.~E.,  2010, \mn@doi
  [Astrophys. J.] {10.1088/0004-637X/725/2/1918}, 725, 1918

\bibitem[\protect\citeauthoryear{Ozel, Psaltis, Narayan  \& Villarreal}{Ozel
  et~al.}{2012}]{Ozel:2012ax}
Ozel F.,  Psaltis D.,  Narayan R.,   Villarreal A.~S.,  2012, \mn@doi
  [Astrophys. J.] {10.1088/0004-637X/757/1/55}, 757, 55

\bibitem[\protect\citeauthoryear{Raidal, Vaskonen  \& Veermäe}{Raidal
  et~al.}{2017}]{Raidal:2017mfl}
Raidal M.,  Vaskonen V.,   Veermäe H.,  2017, \mn@doi [JCAP]
  {10.1088/1475-7516/2017/09/037}, 1709, 037

\bibitem[\protect\citeauthoryear{Raidal, Spethmann, Vaskonen  \&
  Veerm\"ae}{Raidal et~al.}{2019}]{Raidal:2018bbj}
Raidal M.,  Spethmann C.,  Vaskonen V.,   Veerm\"ae H.,  2019, \mn@doi [JCAP]
  {10.1088/1475-7516/2019/02/018}, 02, 018

\bibitem[\protect\citeauthoryear{Regimbau}{Regimbau}{2011}]{Regimbau:2011rp}
Regimbau T.,  2011, \mn@doi [Res. Astron. Astrophys.]
  {10.1088/1674-4527/11/4/001}, 11, 369

\bibitem[\protect\citeauthoryear{Rosado}{Rosado}{2011}]{Rosado:2011kv}
Rosado P.~A.,  2011, \mn@doi [Phys. Rev. D] {10.1103/PhysRevD.84.084004}, 84,
  084004

\bibitem[\protect\citeauthoryear{Sasaki, Suyama, Tanaka  \& Yokoyama}{Sasaki
  et~al.}{2016}]{Sasaki:2016jop}
Sasaki M.,  Suyama T.,  Tanaka T.,   Yokoyama S.,  2016, \mn@doi [Phys. Rev.
  Lett.] {10.1103/PhysRevLett.117.061101}, 117, 061101

\bibitem[\protect\citeauthoryear{Shandera, Jeong  \& Gebhardt}{Shandera
  et~al.}{2018}]{Shandera:2018xkn}
Shandera S.,  Jeong D.,   Gebhardt H. S.~G.,  2018, \mn@doi [Phys. Rev. Lett.]
  {10.1103/PhysRevLett.120.241102}, 120, 241102

\bibitem[\protect\citeauthoryear{Smyth, Profumo, English, Jeltema, McKinnon  \&
  Guhathakurta}{Smyth et~al.}{2020}]{Smyth:2019whb}
Smyth N.,  Profumo S.,  English S.,  Jeltema T.,  McKinnon K.,   Guhathakurta
  P.,  2020, \mn@doi [Phys. Rev. D] {10.1103/PhysRevD.101.063005}, 101, 063005

\bibitem[\protect\citeauthoryear{Soares-Santos et~al.}{Soares-Santos
  et~al.}{2017}]{Soares-Santos:2017lru}
Soares-Santos M.,  et~al., 2017, \mn@doi [Astrophys. J. Lett.]
  {10.3847/2041-8213/aa9059}, 848, L16

\bibitem[\protect\citeauthoryear{Sugiyama, Kurita  \& Takada}{Sugiyama
  et~al.}{2020}]{Sugiyama:2019dgt}
Sugiyama S.,  Kurita T.,   Takada M.,  2020, \mn@doi [Mon. Not. Roy. Astron.
  Soc.] {10.1093/mnras/staa407}, 493, 3632

\bibitem[\protect\citeauthoryear{Tisserand et~al.}{Tisserand
  et~al.}{2007}]{Tisserand:2006zx}
Tisserand P.,  et~al., 2007, \mn@doi [Astron. Astrophys.]
  {10.1051/0004-6361:20066017}, 469, 387

\bibitem[\protect\citeauthoryear{Vaskonen \& Veermäe}{Vaskonen \&
  Veermäe}{2020}]{Vaskonen:2019jpv}
Vaskonen V.,  Veermäe H.,  2020, \mn@doi [Phys. Rev. D]
  {10.1103/PhysRevD.101.043015}, 101, 043015

\bibitem[\protect\citeauthoryear{Wang, Wang, Huang  \& Li}{Wang
  et~al.}{2018}]{Wang:2016ana}
Wang S.,  Wang Y.-F.,  Huang Q.-G.,   Li T. G.~F.,  2018, \mn@doi [Phys. Rev.
  Lett.] {10.1103/PhysRevLett.120.191102}, 120, 191102

\bibitem[\protect\citeauthoryear{Wang, Terada  \& Kohri}{Wang
  et~al.}{2019}]{Wang:2019kaf}
Wang S.,  Terada T.,   Kohri K.,  2019, \mn@doi [Phys. Rev.]
  {10.1103/PhysRevD.99.103531}, D99, 103531

\bibitem[\protect\citeauthoryear{Wyrzykowski et~al.}{Wyrzykowski
  et~al.}{2011}]{Wyrzykowski:2011tr}
Wyrzykowski L.,  et~al., 2011, \mn@doi [Mon. Not. Roy. Astron. Soc.]
  {10.1111/j.1365-2966.2011.19243.x}, 416, 2949

\bibitem[\protect\citeauthoryear{{Zel'dovich} \& {Novikov}}{{Zel'dovich} \&
  {Novikov}}{1967}]{1967SvA....10..602Z}
{Zel'dovich} Y.~B.,  {Novikov} I.~D.,  1967, \sovast, \href
  {https://ui.adsabs.harvard.edu/abs/1967SvA....10..602Z} {10, 602}

\bibitem[\protect\citeauthoryear{Zhu, Howell, Regimbau, Blair  \& Zhu}{Zhu
  et~al.}{2011}]{Zhu:2011bd}
Zhu X.-J.,  Howell E.,  Regimbau T.,  Blair D.,   Zhu Z.-H.,  2011, \mn@doi
  [Astrophys. J.] {10.1088/0004-637X/739/2/86}, 739, 86

\bibitem[\protect\citeauthoryear{Zumalacarregui \& Seljak}{Zumalacarregui \&
  Seljak}{2018}]{Zumalacarregui:2017qqd}
Zumalacarregui M.,  Seljak U.,  2018, \mn@doi [Phys. Rev. Lett.]
  {10.1103/PhysRevLett.121.141101}, 121, 141101

\bibitem[\protect\citeauthoryear{de Lavallaz \& Fairbairn}{de~Lavallaz \&
  Fairbairn}{2010}]{deLavallaz:2010wp}
de Lavallaz A.,  Fairbairn M.,  2010, \mn@doi [Phys. Rev. D]
  {10.1103/PhysRevD.81.123521}, 81, 123521

\makeatother
\end{thebibliography}

\bsp
\label{lastpage}
\end{document}